\documentclass[acmsmall,screen]{acmart}

\usepackage[utf8]{inputenc} 
\usepackage[T1]{fontenc}    
\usepackage{hyperref}       
\usepackage{url}            
\usepackage{booktabs}       
\usepackage{amsfonts}       
\usepackage{nicefrac}       
\usepackage{microtype}      
\usepackage{xcolor}         

\usepackage{graphicx}
\usepackage{pifont}
\usepackage{lipsum}  

\usepackage{algorithm}
\usepackage{algpseudocode}
\usepackage{subcaption}
\usepackage{multirow}

\newcommand{\tick}{\textcolor{green}{\ding{51}}}
\newcommand{\cross}{\textcolor{red}{\ding{55}}}

\newlength{\myMheight}
\usepackage{array}
\newcolumntype{H}{>{\setbox0=\hbox\bgroup}c<{\egroup}@{}}

\title{Kernel-FFI: Transparent Foreign Function Interfaces for Interactive Notebooks}

\author{Hebi Li}
\email{lihebi.com@gmail.com}
\affiliation{%
  \institution{CodePod Inc.}
  \city{Santa Clara}
  \state{California}
  \country{USA}
}

\author{Forrest Sheng Bao}
\email{forrest.bao@gmail.com}
\affiliation{%
  \institution{CodePod Inc.}
  \city{Santa Clara}
  \state{California}
  \country{USA}
}

\author{Qi Xiao}
\email{xiaoqiqiduo@gmail.com}
\affiliation{%
 \institution{LinkedIn}
 \city{Sunnyvale}
 \state{California}
 \country{USA}
}

\author{Jin Tian}
\email{Jin.Tian@mbzuai.ac.ae}
\affiliation{%
  \institution{MBZUAI}
  \city{Dubai}
  \state{United Arab Emirates}
  \country{UAE}
}

\begin{document}

\begin{abstract}
Foreign Function Interfaces (FFIs) are essential for enabling interoperability between programming languages, yet existing FFI solutions are ill-suited for the dynamic, interactive workflows prevalent in modern notebook environments such as Jupyter. Current approaches require extensive manual configuration, introduce significant boilerplate, and often lack support for recursive calls and object-oriented programming (OOP) constructs—features critical for productive, multi-language development.

We present Kernel-FFI, a transparent, language-agnostic framework that enables seamless cross-language function calls and object manipulation within interactive notebooks. Kernel-FFI employs source-level transformation to automatically rewrite cross-language invocations, eliminating the need for manual bindings or boilerplate. Kernel-FFI provides robust support for OOP by enabling foreign object referencing and automatic resource management across language boundaries. Furthermore, to address the blocking nature of Jupyter kernels and support recursive and asynchronous foreign calls, we introduce a novel side-channel communication mechanism.
Our tool will be open-sourced and available at \url{https://codepod.io/docs/kernel-ffi}.
\end{abstract}

\maketitle

\section{Introduction}
\label{sec:introduction}

Modern software development increasingly relies on multi-language architectures to leverage the unique strengths of different programming languages. Python excels in data science and machine learning through its rich ecosystem of libraries such as TensorFlow and PyTorch, while JavaScript dominates web development with frameworks like React and Node.js. Similarly, C++ provides low-level control for performance-critical applications, and Lisp offers unparalleled flexibility for building Domain Specific Languages (DSLs). However, enabling seamless interaction between these languages remains a significant challenge, particularly in interactive development environments. Foreign Function Interfaces (FFIs) are the traditional approach for cross-language communication, allowing functions written in one language to be invoked from another.

Interactive development environments have become essential tools for modern software development, particularly in domains requiring rapid prototyping and exploratory programming. Jupyter Notebooks~\cite{jupyter} exemplify this paradigm, offering a browser-based environment that integrates code, text, and execution results in a literate programming model. By supporting incremental evaluation through a Read-Eval-Print-Loop (REPL) model~\cite{mccarthy1965lisp,van2020principled}, Jupyter has achieved widespread adoption in data science, machine learning, and education~\cite{wang2020better,randles2017using,perkel2018jupyter,pimentel2019large}. The 2024 GitHub Octoverse survey reports over 1.5 million GitHub repositories with Jupyter Notebooks, marking a 92\% year-over-year increase~\cite{github_octoverse_2024}.

Despite supporting multiple language kernels, Jupyter kernels operate in isolation, with no native support for cross-language communication. This limitation significantly constrains the potential for multi-language development within the Jupyter ecosystem. The research landscape for Jupyter-based FFIs remains largely unexplored, presenting a significant gap in the interactive programming literature.

While numerous FFI frameworks exist, they primarily focus on external C library bindings or client-server communications, often requiring extensive manual configuration and boilerplate code. These approaches are ill-suited for the dynamic, exploratory nature of modern interactive programming environments.

Developing an FFI framework for Jupyter's interactive programming paradigm presents several fundamental challenges. First, existing FFIs require substantial manual setup, including boilerplate code generation, memory management, separate library compilation, and complex data serialization. In Jupyter's interactive exploratory programming paradigm, where developers incrementally write and modify code—adding and removing functions, changing function signatures, and redefining classes—the overhead of traditional FFI setup becomes prohibitive.

Second, the blocking nature of Jupyter kernels complicates recursive foreign calls. Naive approaches that send foreign code directly to other kernels introduce blocking behavior, preventing the seamless integration required for interactive development workflows.

Third, Object-Oriented Programming (OOP) is ubiquitous in modern programming languages and widely used in Jupyter code cells, yet existing FFIs often lack support for these paradigms. The inability to instantiate foreign classes, call methods on foreign objects, or maintain object state across language boundaries severely limits the practical utility of current FFI solutions.

In response to these challenges, we present \textbf{Kernel-FFI}, a novel framework designed to enable seamless cross-language interoperability within Jupyter Notebooks. Kernel-FFI addresses the limitations of existing FFIs through three key innovations:

\begin{itemize}
    \item \textbf{Transparent Source-Level Transformation:} Kernel-FFI automatically rewrites cross-language invocations at the source level, allowing developers to call functions in other languages as if they were native. This eliminates the need for manual bindings and boilerplate code, significantly reducing development effort and enabling truly interactive multi-language development.
    
    \item \textbf{Foreign Object Referencing:} Kernel-FFI provides comprehensive support for OOP paradigms by enabling developers to create object instances in one language and reference them in another. This enables seamless interaction with object-oriented code across languages, simplifying the management of stateful libraries and complex data structures.

    \item \textbf{Non-blocking Recursive Calls:} Kernel-FFI introduces side-channel communication to enable non-blocking recursive calls between kernels. This allows developers to write interactive code that can seamlessly switch between languages without introducing latency or blocking behavior, supporting complex multi-language workflows.
\end{itemize}

Kernel-FFI is designed to be language-agnostic, supporting a wide range of programming languages commonly used in data science, web development, and system programming. The framework leverages Jupyter's existing messaging protocol to facilitate communication between kernels, ensuring compatibility with the Jupyter ecosystem while extending its capabilities. By providing a transparent and efficient interface for cross-language function calls, Kernel-FFI enables developers to leverage the strengths of different languages without the burden of complex setup or manual bindings.
Our tool is open-sourced and available at \url{https://codepod.io/docs/kernel-ffi}.

\section{Approach}

In this section, we first formulate the problem of cross-language function calls in Jupyter Notebooks, and then describe the approach to solve the problem.

\subsection{Problem Formulation}

The problem of cross-language interoperability in Jupyter Notebooks is to enable programs written in a source language $L_s$ to seamlessly use functions, classes, and variables defined in a target language $L_t$, in a way that is transparent, interactive, and robust to the dynamic nature of notebook development.

To achieve this, we require a system that:
\begin{itemize}
    \item \textbf{Dynamically discovers and maintains a registry} of programming constructs (functions, classes, variables) defined in each kernel, reflecting the current state of the notebook as code is added, modified, or removed.
    \item \textbf{Automatically transforms} cross-language usages in the source program $P_s$ into a sequence of intermediate representations (IRs) that encode the intent of the foreign call. These IRs must support function calls, variable references, method calls on foreign objects, class instantiation, and object deletion.
    \item \textbf{Serializes and deserializes} arguments and return values across language boundaries, including the ability to reference and manage foreign objects via a global variable/object store, preserving object identity and state.
    \item \textbf{Ensures semantic preservation, type safety, and state consistency} across kernels, so that the transformed program $P_s'$ behaves as if all constructs were native to $L_s$.
\end{itemize}

Formally, for each cross-language usage in $P_s$, the system must:
\begin{enumerate}
    \item Identify the construct in the registry and generate an appropriate IR (as defined in Section~\ref{sec:ir} and the formal theory).
    \item Serialize arguments (including object references) into a language-agnostic format.
    \item Transmit the IR to the target kernel, where it is decoded, executed, and the result is re-encoded.
    \item Deserialize the result in the source kernel, reconstructing native values or proxy objects as needed.
    \item Update the global object store to maintain consistent state and enable further cross-language interactions.
\end{enumerate}

The system must guarantee that, for any sequence of cross-language operations, the observable behavior of $P_s'$ is semantically equivalent to the intended behavior of $P_s$, with type correctness and consistent state across all participating kernels.

\subsection{Architecture Overview}

\begin{figure}
  \centering
  \includegraphics[width=\linewidth]{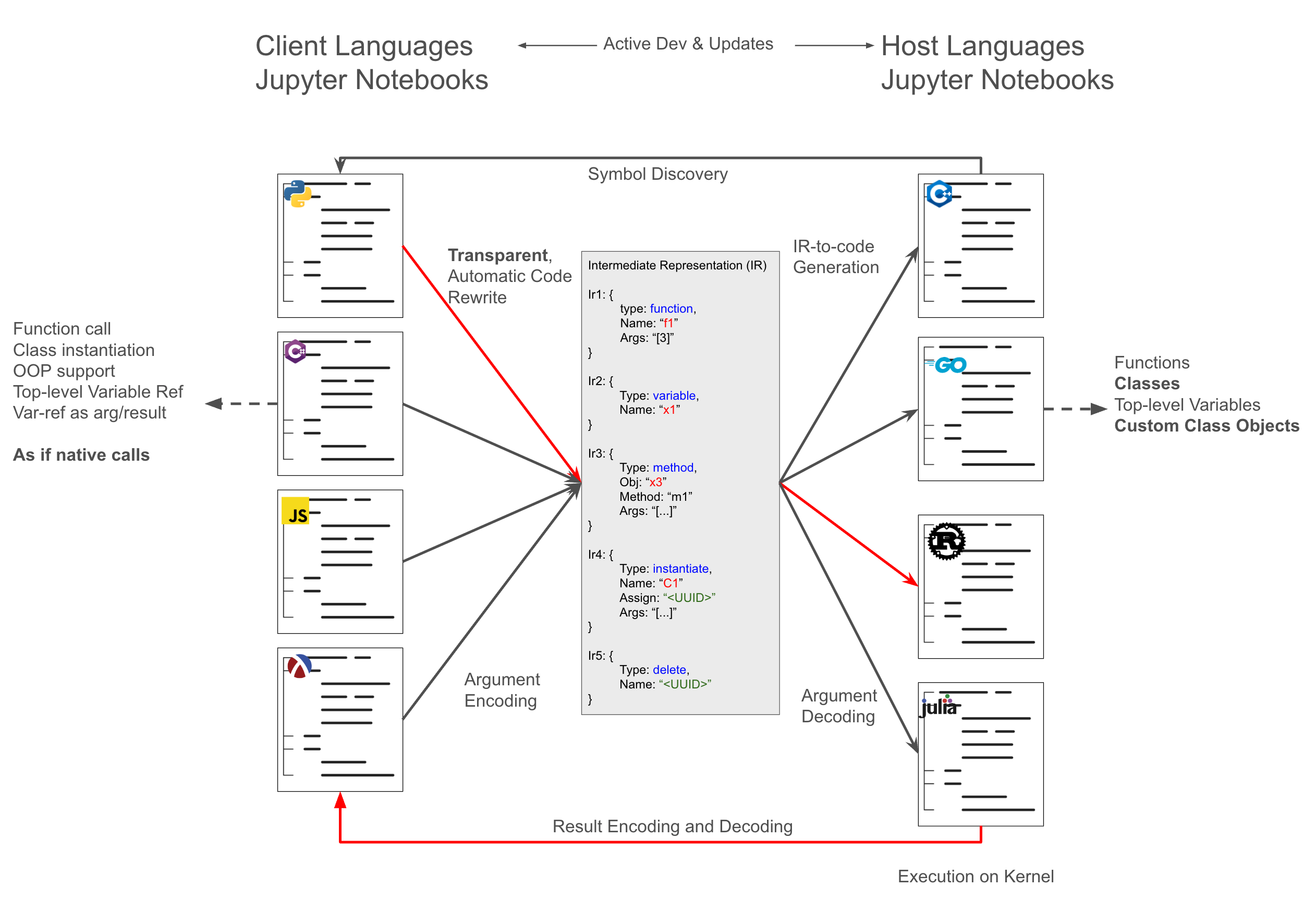}
  \caption{Overall architecture of the Kernel-FFI framework.}
  \label{fig:kernel-ffi-intro}
\end{figure}

Figure \ref{fig:kernel-ffi-intro} shows the overall architecture of Kernel-FFI. We assume that users are developing multiple Jupyter notebooks of different programming languages simultaneously. The host language notebooks (right column) contain programming constructs of interest, such as functions, classes, and variables. The client language notebooks (left column) contain code that wishes to call functions defined in the host language notebooks.

Our system parses and analyzes the host language notebooks to identify programming constructs of interest, such as functions, classes, and variables. These constructs are then registered in a discoverable registry that tracks which functions are available for foreign calls.

The client notebooks can reference these programming constructs. Our system analyzes the client notebooks to identify usage patterns of the programming constructs. Each usage pattern is transformed into a corresponding intermediate representation (IR). Arguments to foreign functions are encoded and embedded in the IR. From the IR, our system generates the corresponding code in the target language. The generated code is sent to the target language kernel for execution. Arguments are decoded from the IR into objects in the host language runtime. Once the target language kernel executes the code, the return value is encoded and sent back to the client language kernel. The return value is then decoded into an object in the client language runtime.

The system is language-agnostic and can support different programming languages. It is designed to be transparent to users working with multiple Jupyter notebooks in different languages. The architecture handles foreign calls across different languages without requiring pairwise mappings between source and target languages. Our approach supports common programming constructs such as functions, classes, and variables that are frequently used in Jupyter notebooks. For example, our approach allows custom class objects to be used in foreign notebook kernels. Additionally, we support recursive foreign calls that are not natively supported by Jupyter kernel's blocking nature.

Our approach is designed to be transparent to users of multiple Jupyter notebooks with different languages. Developers can use the programming constructs of the host notebooks as if they were native to the client notebooks. The entire code transformation is transparent to users.

As shown in Figure~\ref{fig:kernel-ffi-intro}, there may be multiple host language notebooks and multiple client language notebooks of different languages. The IR-based source-level transformation system matches source and target languages without needing to develop pairwise mappings between languages.

\subsection{Approach Overview}

In this section, we show concrete examples of how the Kernel-FFI framework works.

\begin{figure}[ht]
    \centering
    \includegraphics[width=\linewidth]{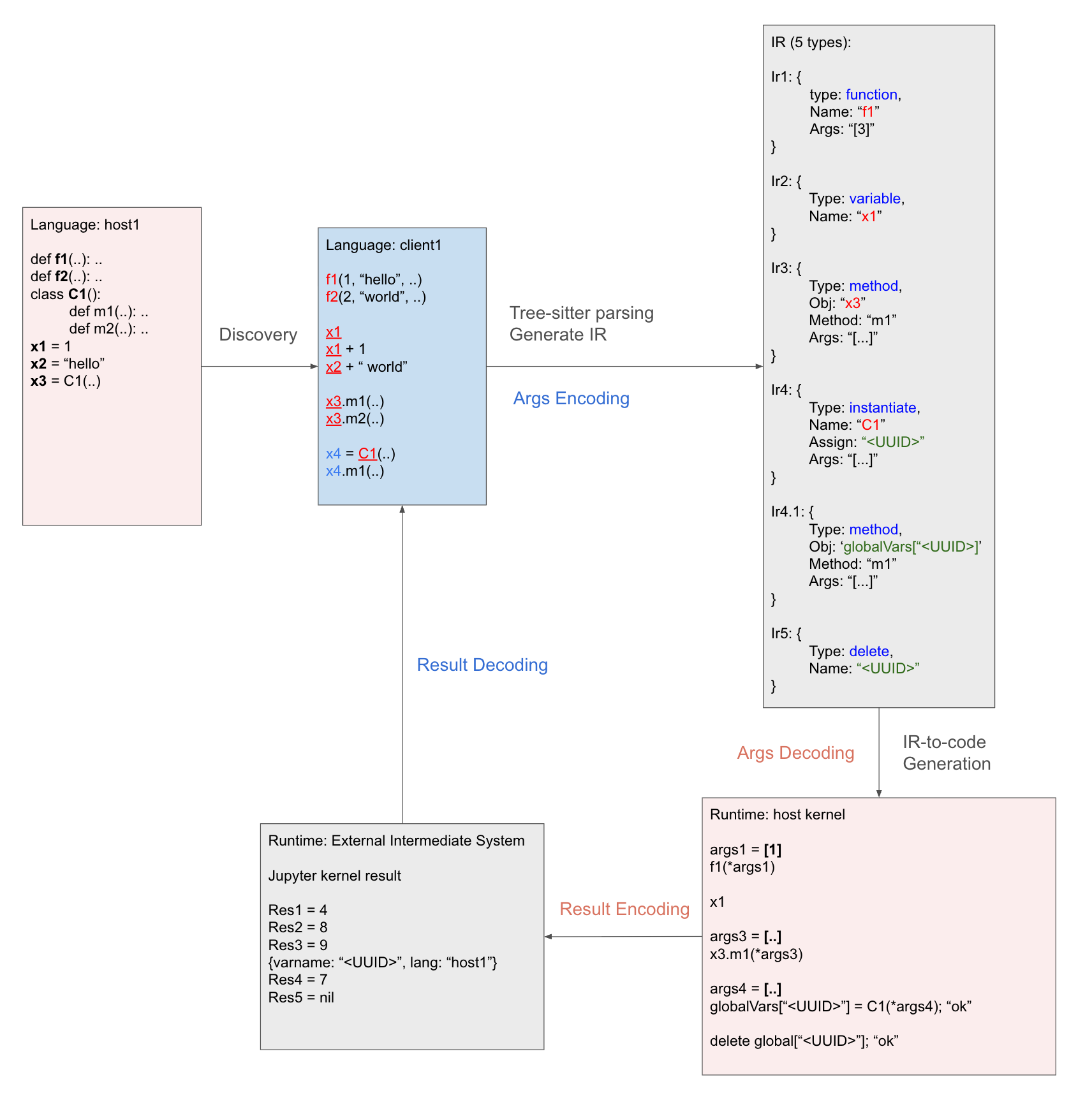}
    \caption{An overall architecture of the Kernel-FFI framework.}
    \label{fig:popl-simple-loop}
\end{figure}

Figure~\ref{fig:popl-simple-loop} shows the overall approach of Kernel-FFI. The host language (host1) uses Python syntax, and the client language (client1) uses TypeScript/JavaScript syntax. We show a simple example where the host program defines functions \texttt{f1} and \texttt{f2}, a class \texttt{C1}, and variables \texttt{x1} and \texttt{x2} (primitive types), and \texttt{x3} (an instance of class \texttt{C1}) in the host runtime. These symbols are parsed with tree-sitter and registered to be discoverable by the client program.

The client program makes use of the host program. For example, it calls the function \texttt{f1} with arguments, refers to remote primitive type variables and uses their values in further computation (e.g., \texttt{x1 + 1}), and calls methods on the custom object \texttt{x3}. Additionally, the client program creates a local variable that is an instance of the remote class \texttt{C1} and calls the method \texttt{m1} on it. The client program is first parsed and matched against the host program constructs.

To execute the foreign calls in the client program, our system generates intermediate representations (IRs) of the foreign calls. We detect usage patterns of the programming constructs and generate corresponding IRs. Each IR is a JSON string that encodes the type and attributes of the foreign call. Based on the remote symbol type and usage pattern, we generate IRs of five types: (1) function calls, (2) variable references, (3) method calls on foreign objects, (4) instantiation of foreign classes and calls to their methods, and (5) deletion of foreign objects (triggered by garbage collection, e.g., when objects are re-declared in the REPL or go out of scope). The IRs are language-agnostic and can be used to generate code in any target language. We discuss the details of the IRs in Section~\ref{sec:ir}.

From the IR, our system generates target code in the host language. The generated code is routed to the appropriate kernel for execution. Arguments are decoded from strings into objects of the corresponding types in the target language. Function and method calls are then executed with the decoded arguments. The generated code is sent to the Jupyter kernel for execution.

Once the Jupyter kernel executes and returns a result, the result is serialized into a JSON string and sent back to the client program. The client program decodes the JSON string into a corresponding object in the client language. The client program can then use this object as if it were a native object of the client language.


\subsection{Identifying the programming constructs of interest}

The first step in our approach is to identify programming constructs that can be called from foreign kernels. We focus on common constructs frequently used in Jupyter notebooks: functions, classes, and variables.

To identify these constructs, we employ tree-sitter parsers to analyze source code in the host kernel. Tree-sitter provides robust parsing capabilities across multiple programming languages, enabling consistent extraction of syntactic information. The parser generates an Abstract Syntax Tree (AST) that we traverse to identify function definitions (both global and class-scoped), class definitions and their methods, and variable declarations and assignments at global scope.

For each identified construct, we maintain a registry that tracks essential information about the programming elements. This includes the construct's name and type (function, class, or variable), its scope and visibility, parameter information for functions and methods, and method definitions and inheritance hierarchy for classes.

The registry serves multiple critical purposes in our system. First, it provides a lookup mechanism to validate foreign calls, ensuring they reference existing constructs. Second, it enables type checking and argument validation before execution. Third, it facilitates proper serialization and deserialization of arguments and return values.

The identification process is dynamic and continuous, reflecting the interactive nature of Jupyter notebooks. As new cells are executed in the host kernel, the registry is updated to include newly defined constructs or remove deleted ones. This ensures that the foreign call interface always reflects the current state of the host kernel's environment.

To handle the dynamic nature of interactive programming, we track modifications to existing constructs. When a function or class is redefined, we update the registry accordingly and ensure that subsequent foreign calls use the latest definition. This is particularly important in notebook environments where code cells can be re-executed in any order. The continuous monitoring and updating of the registry enables our system to maintain consistency between host and client kernels, even as the codebase evolves during interactive development.

\subsection{Generating IRs}
\label{sec:ir}

\begin{figure}[ht]
    \centering
    \includegraphics[width=\linewidth]{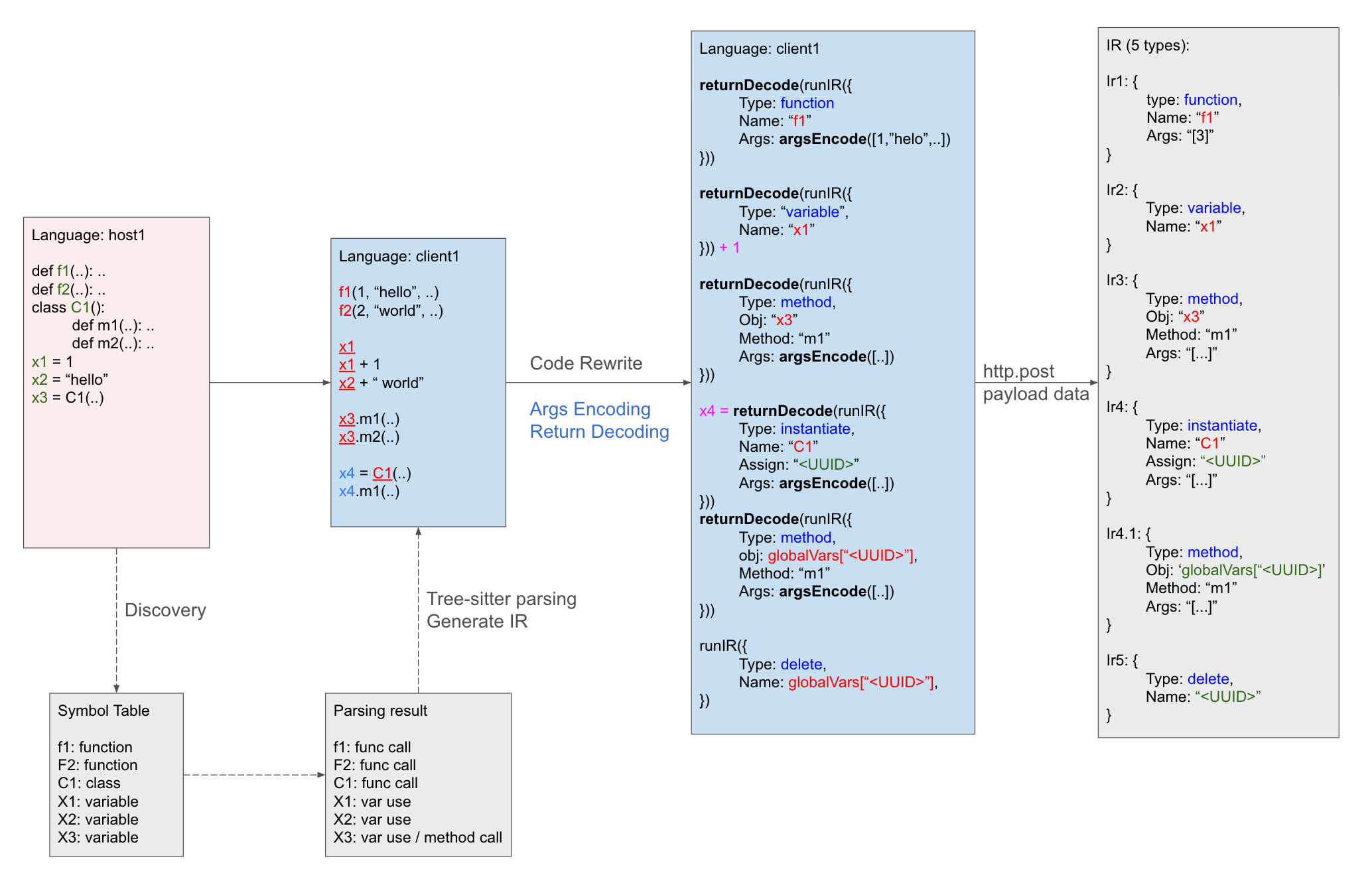}
    \caption{Generating IRs from the source program.}
    \label{fig:popl-clientRewrites}
\end{figure}

The first step in enabling transparent FFI is to generate Intermediate Representations (IRs) from the source program. The IR generation process involves identifying programming constructs of interest and transforming them into a standardized format that can be understood by any target language kernel.

When generating IRs, we need to handle different types of programming constructs:

\begin{itemize}
  \item For function calls to the host program, the IR contains three key fields: \texttt{type} which is set to \texttt{function}, \texttt{name} which specifies the function name to be called, and \texttt{args} which contains the serialized list of arguments. For example, a function call \texttt{calculate(x, y)} would be represented as \texttt{\{"type": "function", "name": "calculate", "args": "[1, 2]"\}}.

  \item For variable references, the IR is simpler, containing just two fields: \texttt{type} set to \texttt{variable} and \texttt{name} indicating the variable name being accessed. This allows direct access to variables defined in the host program. For instance, accessing a variable \texttt{result} would generate \texttt{\{"type": "variable", "name": "result"\}}.

  \item Method calls on foreign objects require additional information in the IR. The \texttt{type} field is set to \texttt{method}, while \texttt{obj} specifies the object instance, \texttt{method} indicates the method name being called, and \texttt{args} contains the serialized method arguments. A method call like \texttt{obj.process(data)} would be represented as \texttt{\{"type": "method", "obj": "obj", "method": "process", "args": '["data"]'\}}.

  \item When instantiating foreign classes, the IR uses the \texttt{type} field set to \texttt{instantiate}, along with \texttt{class} to specify the class name and \texttt{args} for the constructor arguments. This enables object creation in the host program that can be referenced from the client program. Creating an instance like \texttt{MyClass(param1, param2)} would generate \texttt{\{"type": "instantiate", "class": "MyClass", "args": "[1, 2]"\}}.

  \item For cleaning up foreign objects, a deletion IR is generated with \texttt{type} set to \texttt{delete} and \texttt{obj} indicating the object to be removed. This ensures proper resource management across language boundaries when objects are no longer needed. Deleting an object \texttt{obj} would produce \texttt{\{"type": "delete", "name": "obj"\}}.
\end{itemize}

The arguments list is obtained by serializing the arguments into a JSON string. Specifically, we put the arguments into a list and then serialize the list into a JSON string. To support remote object reference for variables and Object-Oriented Programming (OOP), we introduce special serialization and deserialization of objects.

The IR generation process is language-agnostic, meaning it can be implemented for any source language that supports these basic programming constructs. The generated IRs serve as a bridge between the source and target languages, enabling uniform handling of cross-language operations.

For argument serialization during IR generation, we employ a specialized encoding scheme that preserves object references and handles complex data structures. This is particularly important for maintaining object identity across language boundaries and supporting object-oriented programming patterns.

The argument encoding/decoding and return value encoding/decoding handle the serialization and deserialization of arguments and return values. This supports language-agnostic intermediate representation and payload transmission between host and client kernels. For simple types, encoding and decoding are straightforward. For example, the integer 1 is encoded as \texttt{"1"} and decoded as \texttt{1}. The string \texttt{"hello"} is encoded as \texttt{"hello"} and decoded as \texttt{"hello"}. For custom data types, we serialize the object into a remote reference and store the object in a global variable store. The encoding and decoding are introduced in more detail in Section~\ref{sec:object-reference} where object reference is supported.

As shown in Figure~\ref{fig:popl-clientRewrites}, the source program is analyzed to identify foreign function calls and other cross-language operations. These operations are then transformed into their corresponding IR representations, which capture all necessary information for execution in the target kernel while maintaining semantic equivalence with the original code.

\subsection{Generating target code from IRs}



\begin{figure}[ht]
    \centering
    \includegraphics[width=\linewidth]{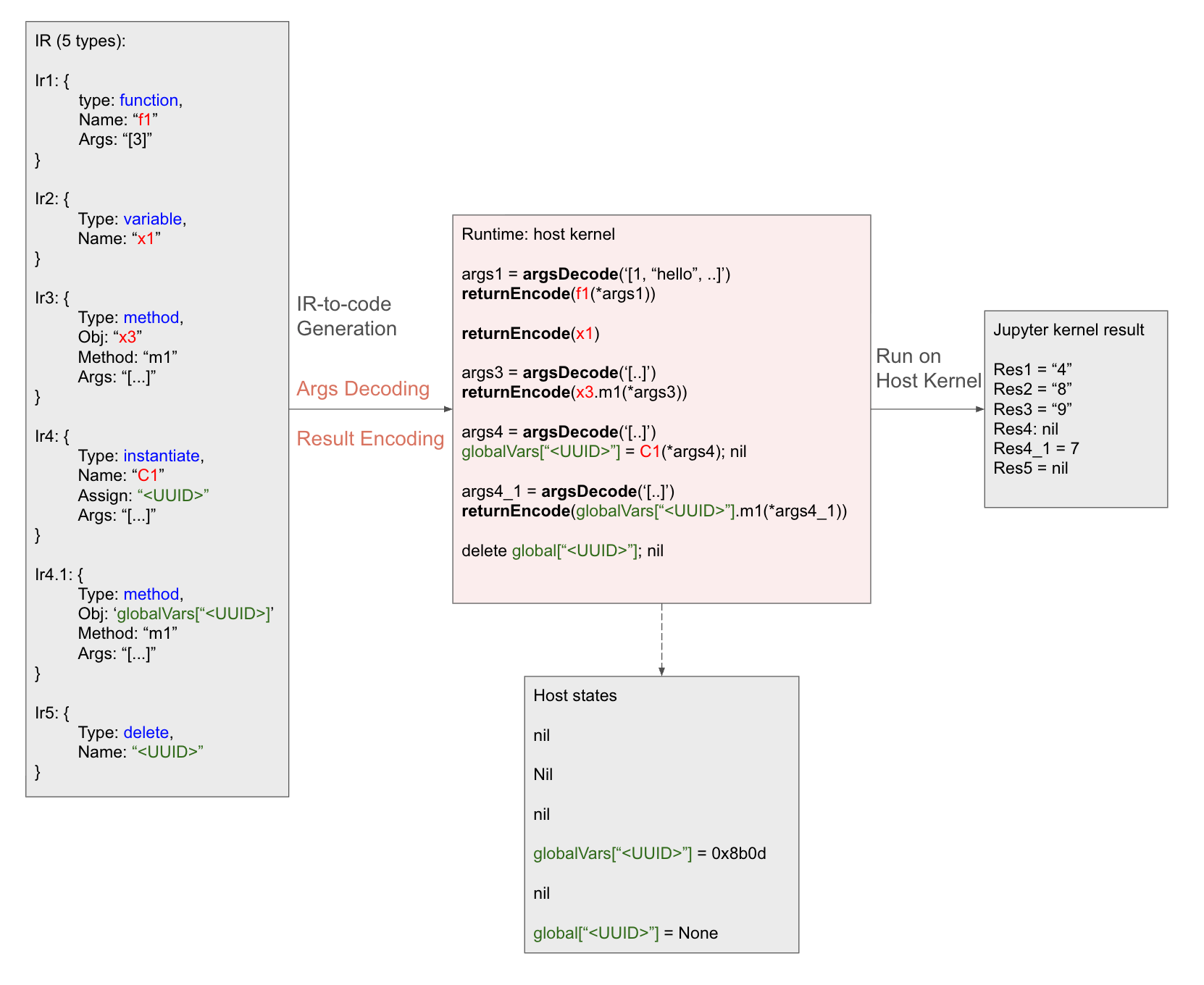}
    \caption{Generating target code from IRs.}
    \label{fig:popl-hostRewrites}
\end{figure}

Once IRs are generated from the source program, they are transformed into executable code in the target language. This transformation process involves converting the IR's abstract representation into concrete syntax that can be executed by the target kernel while handling argument decoding and return value encoding.

The IR-to-code transformation handles different IR types as follows:

\begin{itemize}
    \item For function calls, the IR is transformed into a function invocation with decoded arguments, and the return value is encoded for transmission back to the source language.
    \item Variable access IRs are converted into direct variable references, with appropriate encoding of the retrieved value.
    \item Method call IRs generate code that invokes the specified method on the target object, with proper argument decoding.
    \item Class instantiation IRs create new object instances, storing them in a global variable store for future reference.
    \item Delete IRs remove object references from the global variable store to manage memory.
\end{itemize}

As illustrated in Figure~\ref{fig:popl-hostRewrites}, the transformation process ensures that all necessary type conversions and object reference management are handled automatically. The generated code includes calls to specialized encoding and decoding functions that maintain semantic consistency across language boundaries.

For example, when transforming a function call IR, the generated code might look like:
\begin{verbatim}
myReturnEncode(functionName(*myArgsDecode("[serializedArgs,..]")))
\end{verbatim}

Here, \texttt{myArgsDecode} deserializes the arguments into their appropriate target language representations, the function is called with these decoded arguments, and \texttt{myReturnEncode} serializes the return value for transmission back to the source language.

The IR-to-code transformation system is designed to be extensible, allowing support for new IR types and language-specific features while maintaining the core principle of transparent cross-language execution. This flexibility enables Kernel-FFI to adapt to different programming paradigms and language features while preserving type safety and semantic correctness.

\subsection{Closing the loop}

After executing the target code, the return value is properly encoded and transmitted back to the source language. This process involves serializing the result into a format that can be safely transmitted across language boundaries and correctly interpreted by the source language.

The return value encoding process handles various data types and structures:

\begin{itemize}
    \item Primitive types (numbers, strings, booleans) are encoded directly into their JSON representations
    \item Complex objects are serialized with their type information preserved
    \item References to objects in the target kernel are encoded as special identifiers
    \item Null or undefined values are handled consistently across languages
\end{itemize}

When the encoded return value reaches the source language, it undergoes a decoding process that reconstructs the appropriate data structures in the source language's type system. This decoding process is the inverse of the encoding operation performed in the target kernel, ensuring that the semantic meaning of the data is preserved across language boundaries.

\subsection{Object Oriented Programming with Foreign Object Reference}
\label{sec:object-reference}



In this section, we describe how to support Object Oriented Programming (OOP) with foreign object reference.

\begin{figure}[ht]
  \centering
  \includegraphics[width=0.8\linewidth]{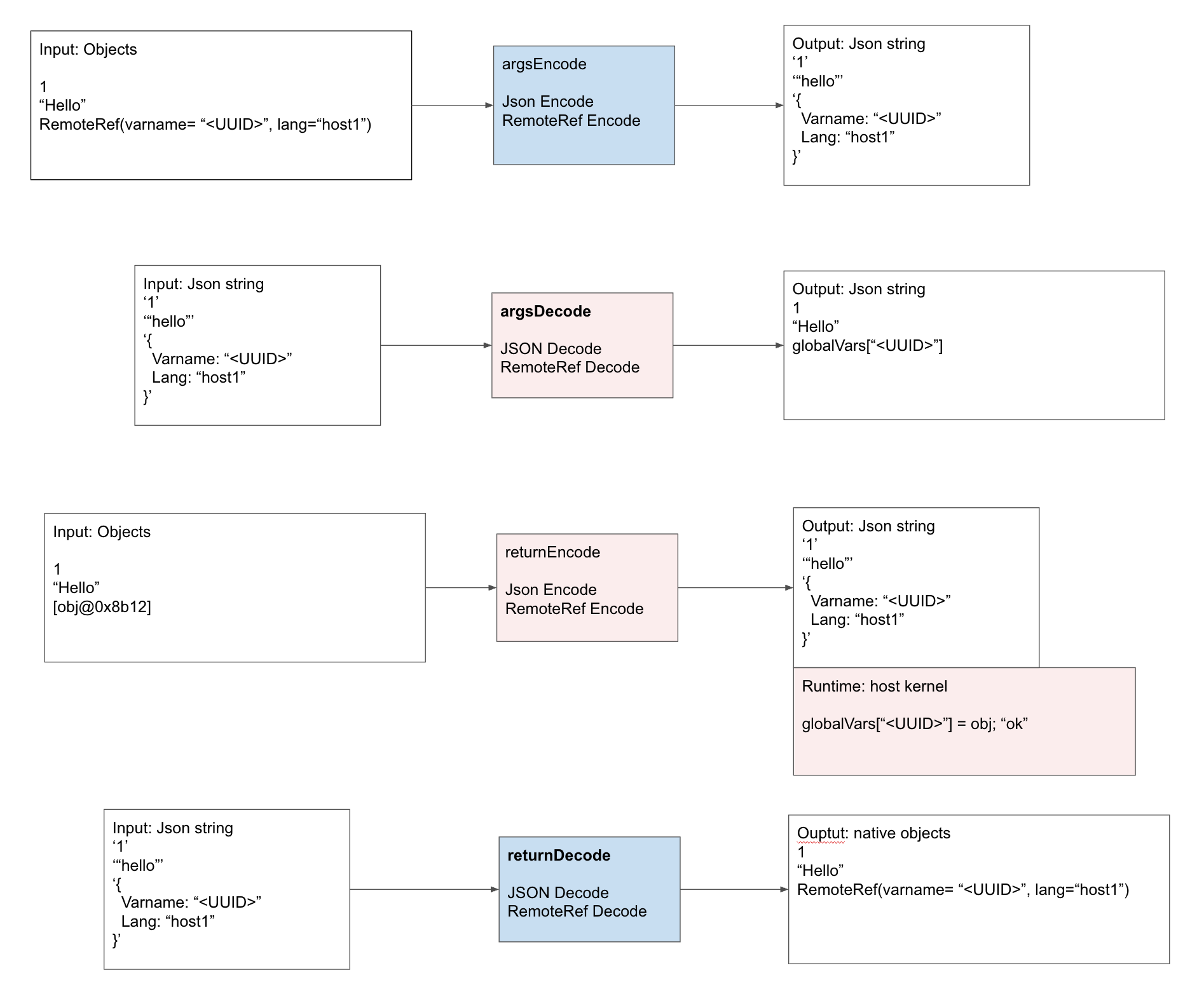}
  \caption{Object Oriented Programming with foreign object reference.}
  \label{fig:popl-encodeDecode}
\end{figure}

Object-Oriented Programming (OOP) is a common paradigm in modern software development, and Kernel-FFI supports seamless interaction with objects across languages. When a class instance is created in a foreign language, Kernel-FFI creates a reference object in the source language's kernel. This reference object maintains a mapping between the original object and its reference, allowing method calls and attribute access to be transparently forwarded to the target language. When the object is no longer needed, Kernel-FFI automatically cleans up the reference object, ensuring that resources are released properly. This approach simplifies the management of object instances and reduces the need for manual bookkeeping or wrapper functions.

\begin{figure}[ht]
  \centering
  \includegraphics[width=\linewidth]{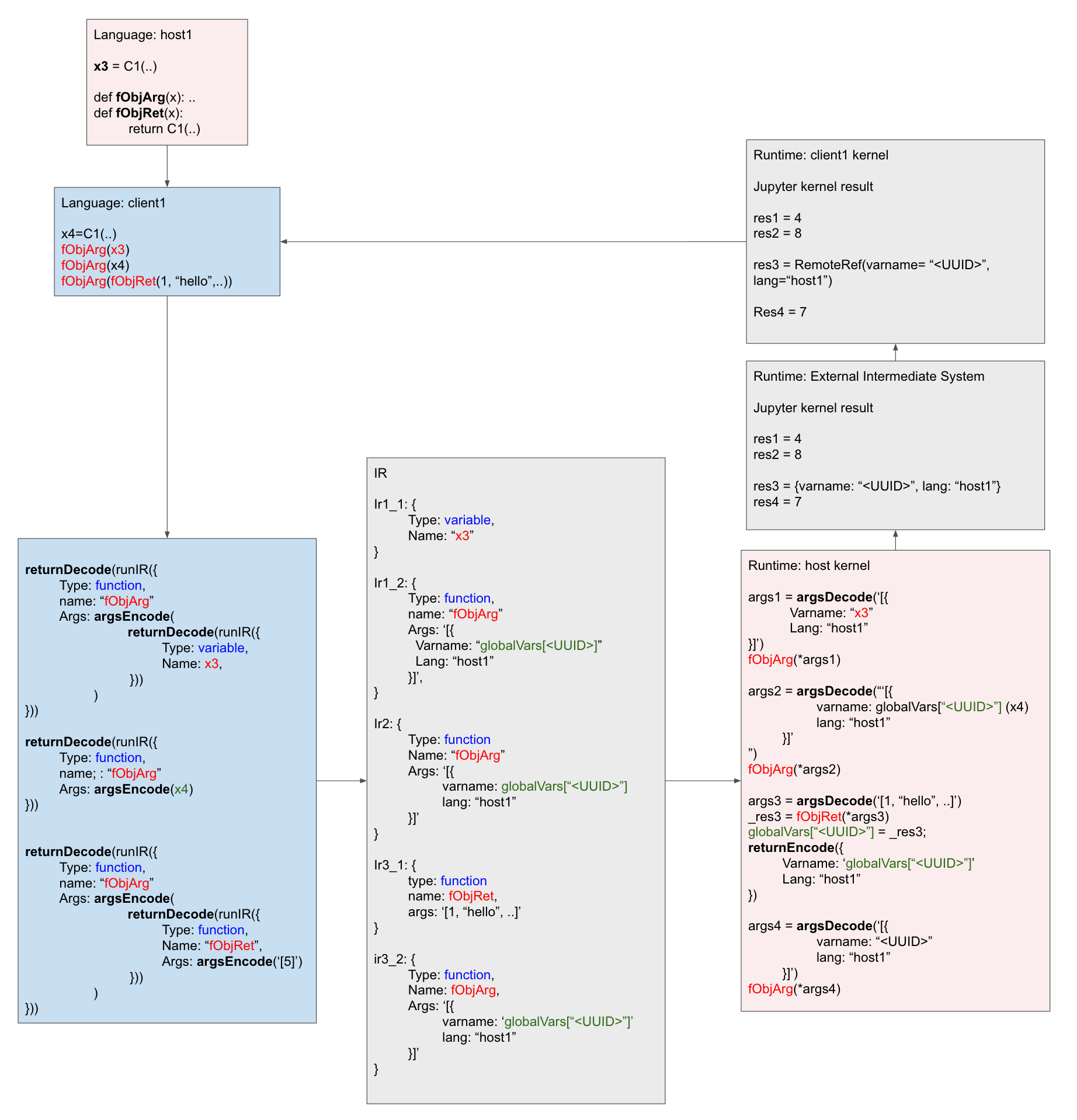}
  \caption{Object Oriented Programming with foreign object reference.}
  \label{fig:popl-objRef}
\end{figure}

Object referencing is handled in the encoding and decoding of arguments and return values. Figure~\ref{fig:popl-encodeDecode} shows the encoding and decoding of arguments and return values. For primitive types and collection types that are serializable by JSON, we can directly encode and decode them into and from JSON strings. For custom type objects, we do not encode the object itself, but instead store a reference to the object in a global store called \texttt{globalVars} with a randomly generated UUID. The reference is encoded as a JSON string with the object's UUID and type information. This UUID serves as a reference to the object in the target kernel, and the type information is used to decode the object in the target kernel for languages that do not support implicit dynamic type casting such as Rust and C++.

\begin{figure}[ht]
  \centering
  \includegraphics[width=0.8\linewidth]{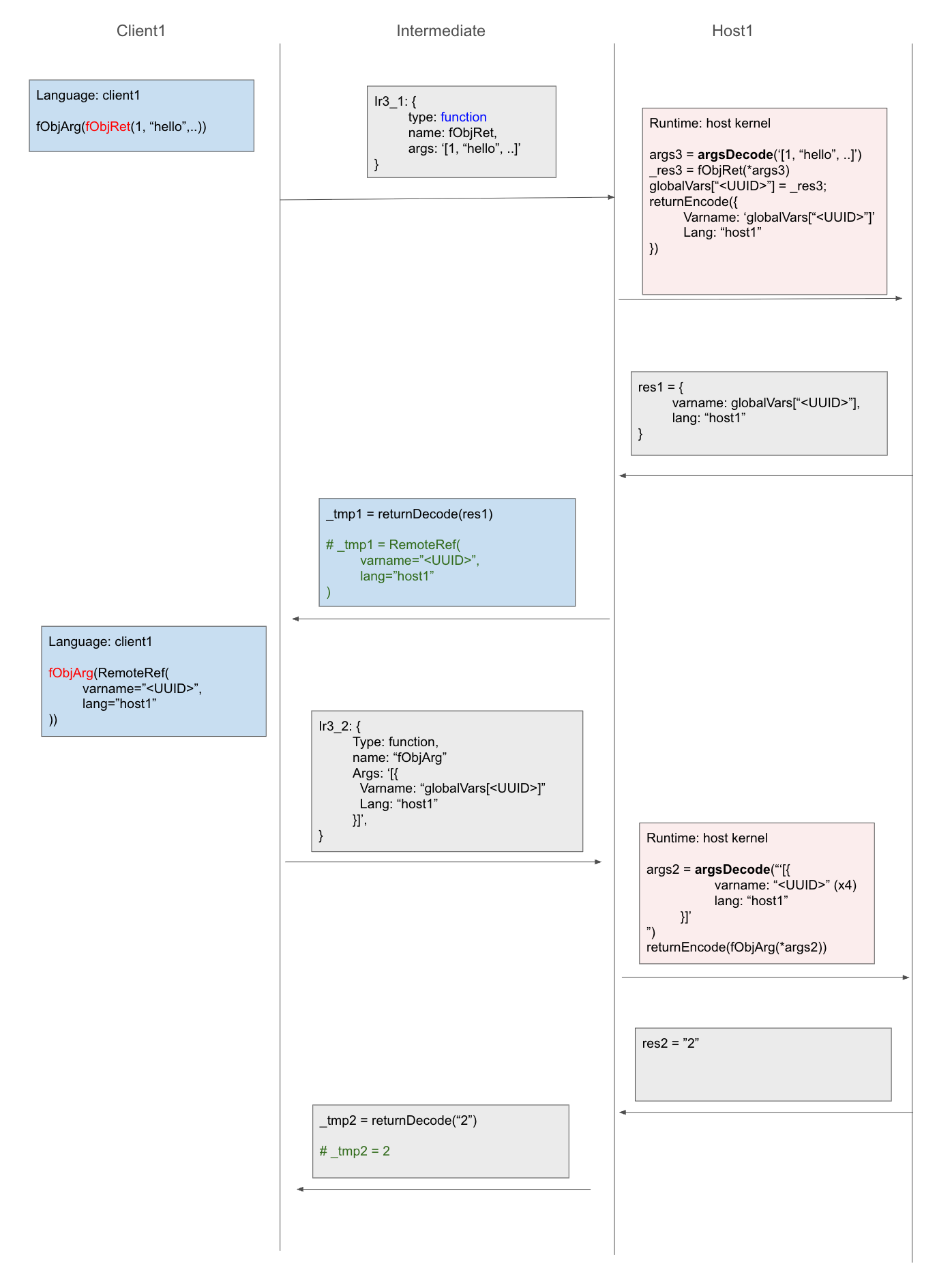}
  \caption{Sequence of operations for object reference.}
  \label{fig:popl-objRefSequence}
\end{figure}

Figure~\ref{fig:popl-objRef} shows a more complex example of object reference. It shows two host functions: \texttt{fObjArg} takes a custom object as an argument, and \texttt{fObjRet} returns a custom object. We show three patterns of using object reference. In the first pattern, the object \texttt{x3} is created in the host language and referenced in the client kernel by calling \texttt{fObjArg} with the remote object. In the second pattern, the object \texttt{x4} is created in the client language and passed to the host language by calling \texttt{fObjArg} with the local object \texttt{x4}. In the third pattern, the client program calls \texttt{fObjRet} to return an object and uses it as an argument of the host function \texttt{fObjArg}.

The corresponding IRs for the three patterns are also shown in Figure~\ref{fig:popl-objRef}. In the first pattern, two IRs are generated: one for variable reference to the remote object \texttt{x3}, and another for a function call to the host function \texttt{fObjArg} with the remote object as the argument. In the second pattern, only one IR is generated: a function call with the local object as the argument. The local object is encoded as a reference to a generated remote object with a random UUID. In the third pattern, two IRs are generated: one for a function call to the host function \texttt{fObjRet} to return an object, and another for a function call to the host function \texttt{fObjArg} with the remote object as the argument.

Figure~\ref{fig:popl-objRefSequence} shows the sequence of operations for the three patterns. In the first pattern, the object \texttt{x3} is created in the host language and referenced in the client kernel by calling \texttt{fObjArg} with the remote object. In the second pattern, the object \texttt{x4} is created in the client language and passed to the host language by calling \texttt{fObjArg} with the local object \texttt{x4}. In the third pattern, the client program calls \texttt{fObjRet} to return an object and uses it as an argument of the host function \texttt{fObjArg}.

Object reference is useful in several scenarios. First, developers often develop classes and work with object instances in notebooks. Foreign object reference allows developers to access object fields and call object methods in notebooks of other languages. Second, object instantiation can be used in more complex scenarios such as loops or nested functions. For example, in a Python loop, a developer can create a new JavaScript Socket instance in each iteration and invoke socket HTTP requests in each iteration.


\subsection{Side Channel Communication for non-Blocking Recursive Calls}

\begin{figure}[ht]
    \centering
    \begin{subfigure}[b]{0.45\linewidth}
        \centering
        \includegraphics[width=\linewidth]{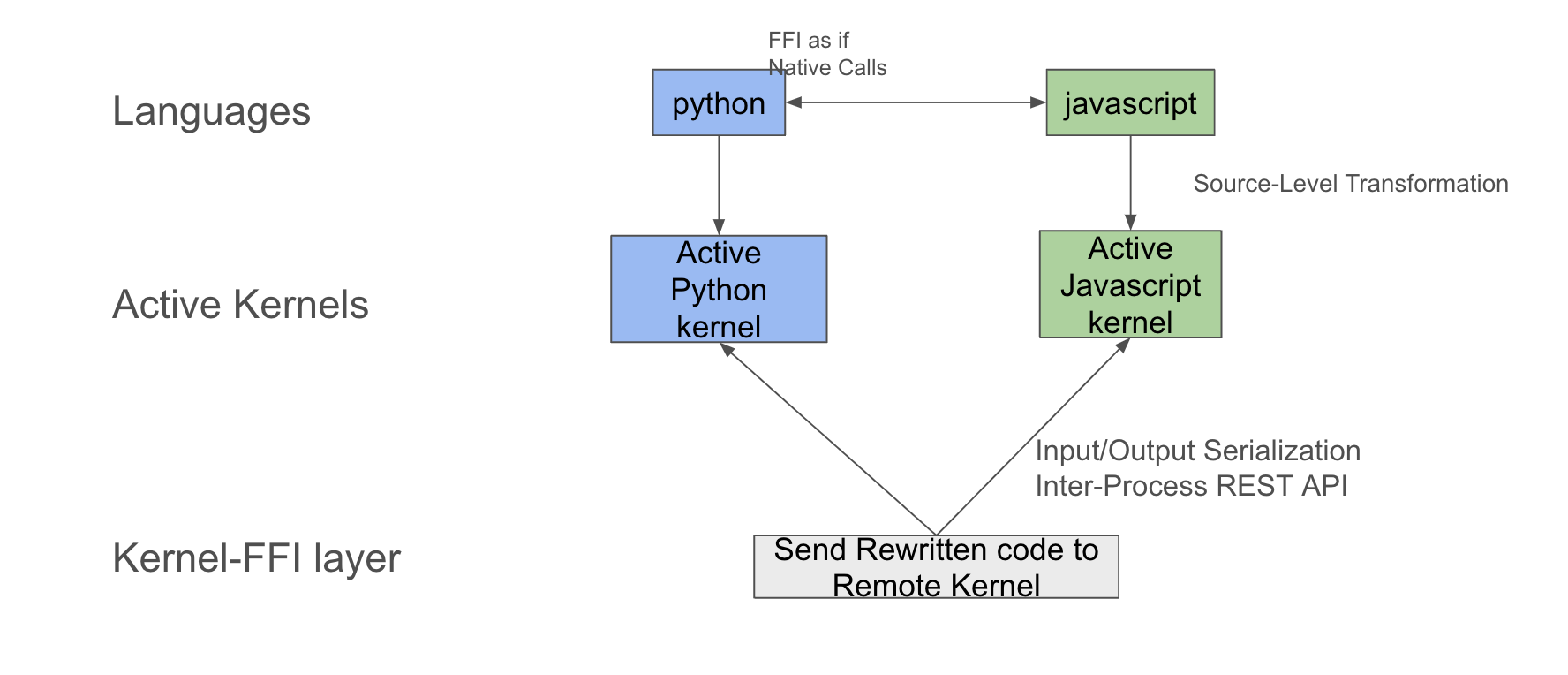}
        \caption{Kernel communication in Kernel-FFI leverages the Jupyter messaging protocol to route code and data between kernels, enabling transparent cross-language function calls.}
        \label{fig:naive-kernel}
    \end{subfigure}
    \hfill
    \begin{subfigure}[b]{0.45\linewidth}
        \centering
        \includegraphics[width=\linewidth]{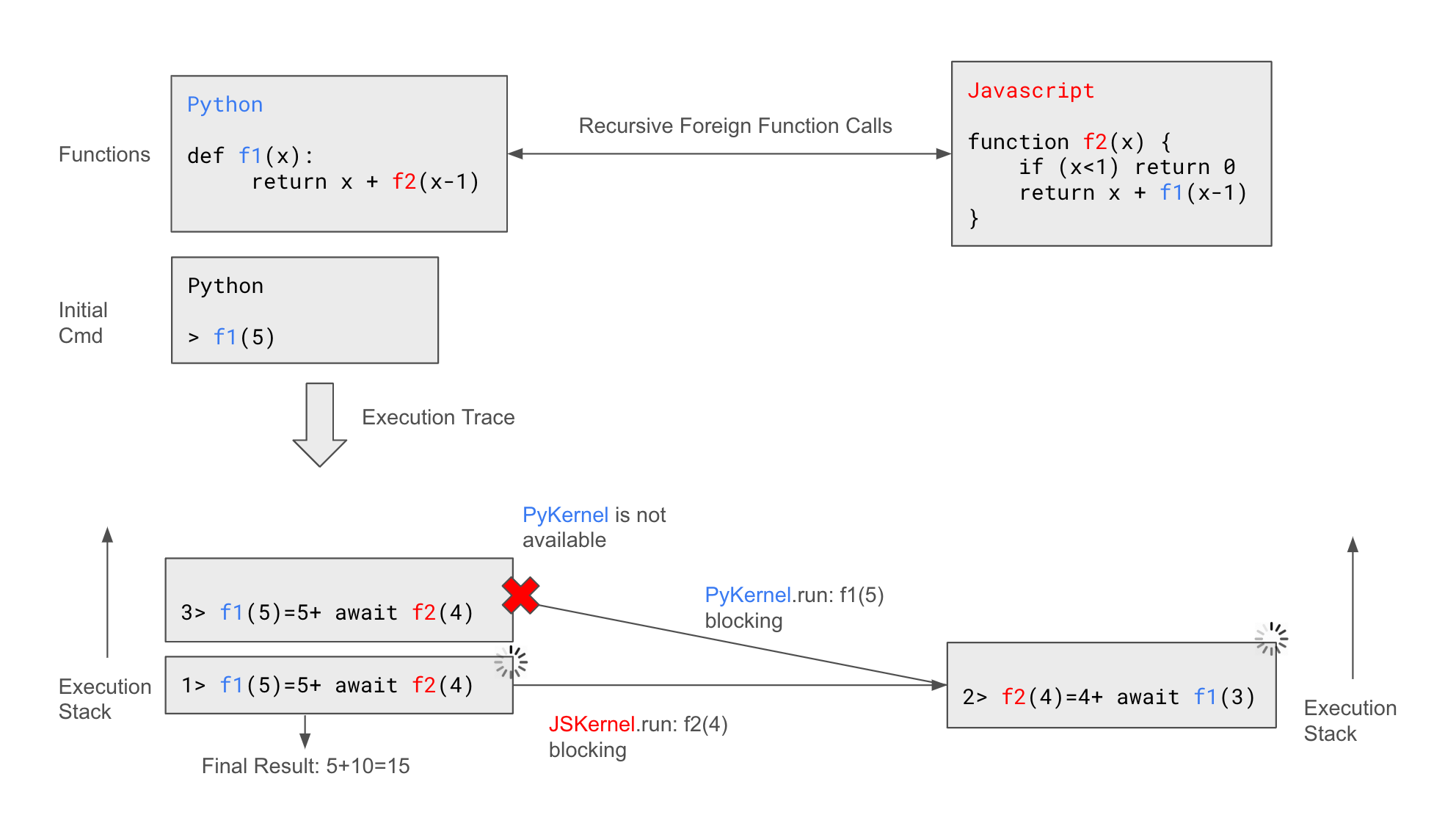}
        \caption{Using the main kernel thread for FFI calls can introduce blocking behavior, preventing recursive foreign calls.}
        \label{fig:kernel-blocking}
    \end{subfigure}
    \caption{Jupyter kernel communication is blocking, which can introduce latency and hinder the performance of recursive or asynchronous calls.}
    \label{fig:jupyter-kernel-blocking}
\end{figure}

To execute the generated target code on the host kernel, one naive approach is to send the generated code to Jupyter's ZMQ wire. However, this approach is blocking. The Jupyter kernel always blocks on one code execution, waiting for results before executing the next code cell. This means that the client kernel waits for the target kernel to execute the code and return the result before proceeding. If the host function being called now calls a remote function in the client kernel, it won't execute because the client kernel is blocking. This synchronous behavior prevents recursive foreign calls. This is illustrated in Figure~\ref{fig:jupyter-kernel-blocking}.

\begin{figure}[ht]
    \centering
    \begin{subfigure}[b]{0.45\linewidth}
        \centering
        \includegraphics[width=\linewidth]{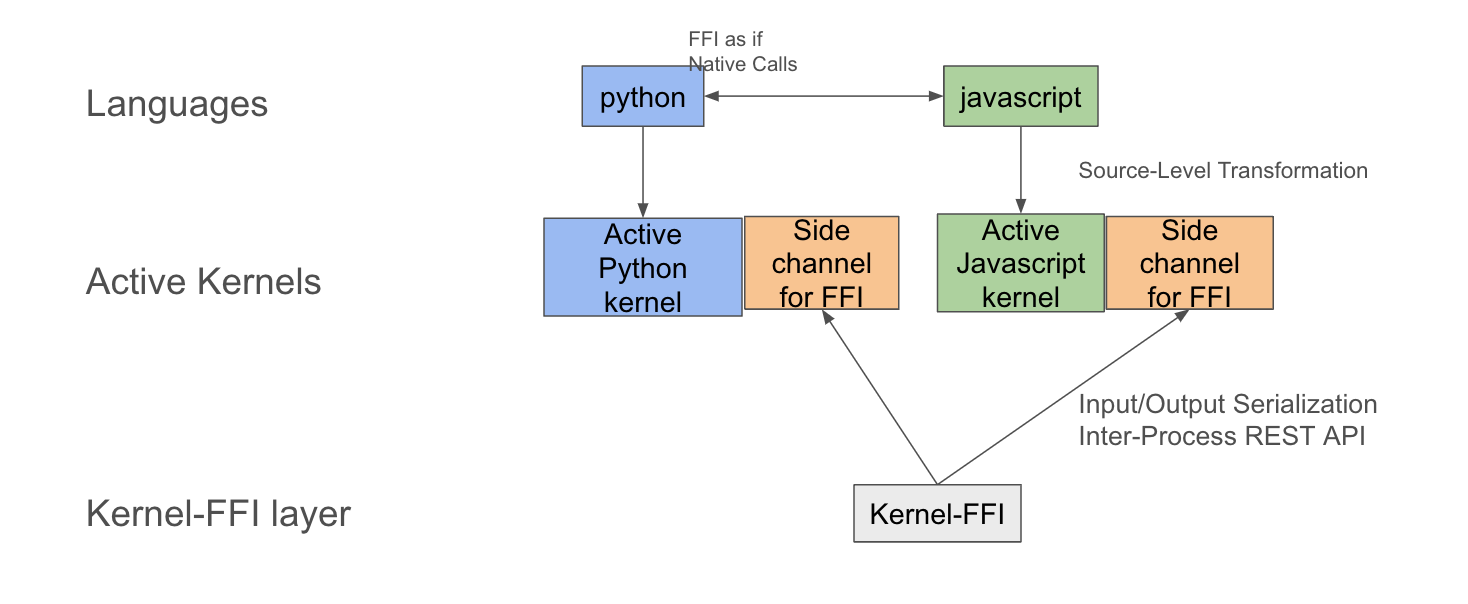}
        \caption{Side-channel communication in Kernel-FFI enables non-blocking recursive calls between kernels, improving performance and responsiveness.}
        \label{fig:side-channel}
    \end{subfigure}
    \hfill
    \begin{subfigure}[b]{0.45\linewidth}
        \centering
        \includegraphics[width=\linewidth]{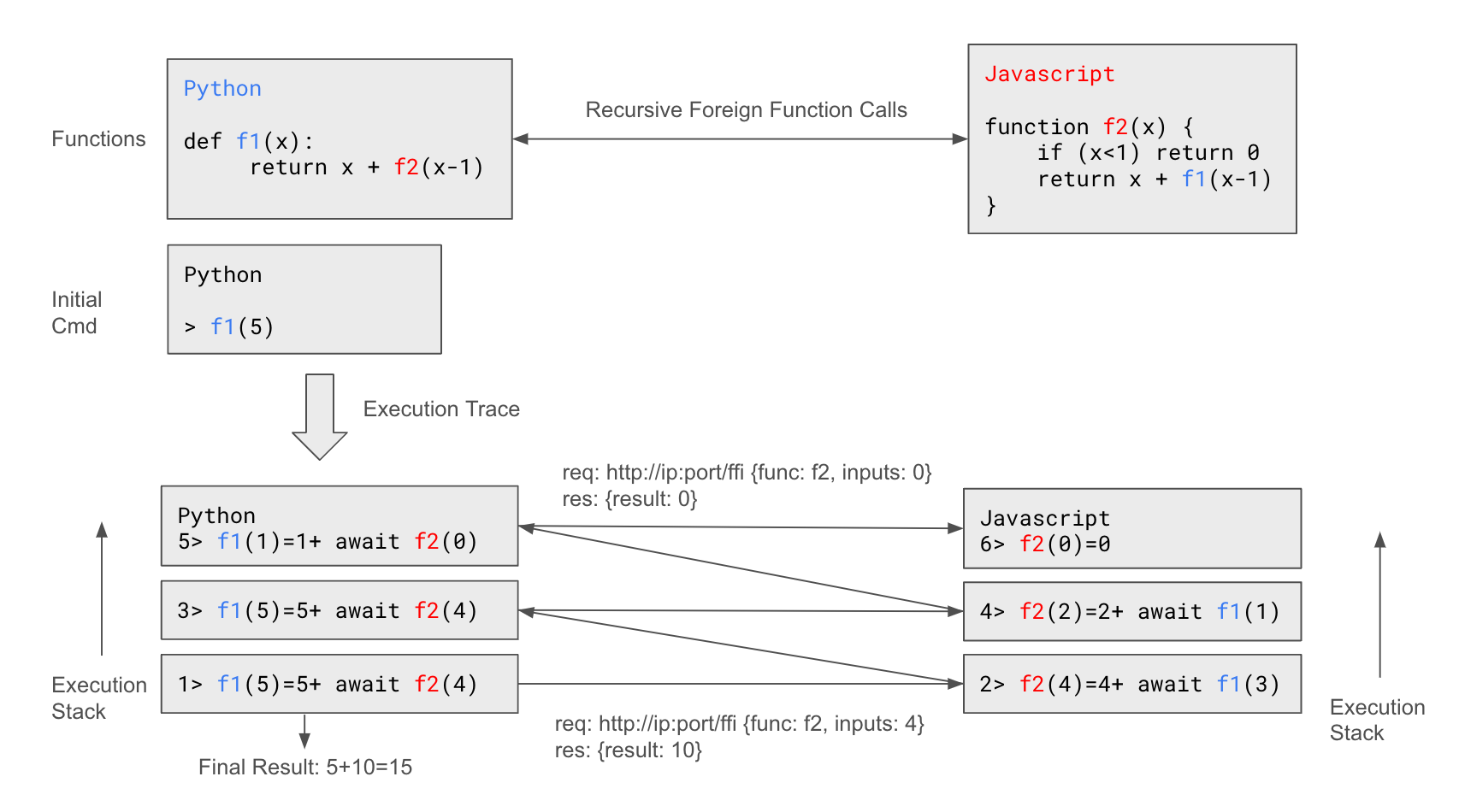}
        \caption{Multi-hop foreign calls in Kernel-FFI enable complex interactions between multiple languages, allowing developers to seamlessly integrate code across different kernels.}
        \label{fig:ffi-stack}
    \end{subfigure}
    \caption{Side-channel communication enables non-blocking recursive calls between kernels, improving performance and responsiveness.}
    \label{fig:side-channel-nonblocking}
\end{figure}

To address this limitation, Kernel-FFI introduces side-channel communication, which allows kernels to communicate asynchronously without blocking the source kernel. Instead of sending the generated target code to the remote Jupyter kernel ZMQ wire, each kernel has a side thread that listens to HTTP requests. When handling an HTTP request, the kernel executes the code and returns the result. Since the HTTP server runs on a separate thread and is non-blocking, this enables recursive foreign function calls between kernels. For example, when kernel A calls a function in kernel B, kernel B can make a call back to kernel A through the side channel without being blocked by the original call. This design allows for complex interactions between kernels while maintaining responsiveness and avoiding deadlocks. Figure~\ref{fig:side-channel} shows the side-channel communication between kernels.

Note that in order to make the side-channel communication work, the language must support \texttt{eval} function that executes the code in the string. In our implementation, the \texttt{eval} function is supported by the dynamic type checked languages including Python, JavaScript, Rust, Ruby and Julia. The other languages such as Rust, Go, C\# and C++ do not support \texttt{eval} function, which we fall back to the naive kernel communication approach. More details are discussed in Section \ref{sec:eval-endpoint}.

\section{Formal Theory and Analysis}


In this section, we formally define the programming language theory for Kernel-FFI. We establish a rigorous framework that captures the semantics of cross-language function calls, object references, and state management across kernels.

\subsubsection{Basic Definitions}

Let $\mathcal{L}$ be the set of supported programming languages, and let $\mathcal{K}$ be the set of Jupyter kernels. For each language $L \in \mathcal{L}$, we define:

\begin{itemize}
    \item $\mathcal{T}_L$: The set of types in language $L$
    \item $\mathcal{V}_L$: The set of values in language $L$
    \item $\mathcal{F}_L$: The set of functions in language $L$
    \item $\mathcal{C}_L$: The set of classes in language $L$
    \item $\mathcal{O}_L$: The set of objects in language $L$
\end{itemize}

\subsubsection{Intermediate Representation (IR)}

An Intermediate Representation (IR) is a JSON structure that encodes cross-language operations. Formally, an IR is defined as:

\begin{align}
IR &= \{(type, name, args) \mid type \in \{function, variable, method, instantiate, delete\}, \\
&\quad name \in \Sigma^*, args \in \Sigma^* \cup \emptyset\}
\end{align}

where $\Sigma^*$ denotes the set of all strings over alphabet $\Sigma$.

For each IR type, we define the structure:
\begin{itemize}
    \item $IR_{function} = \{(type: "function", name: f, args: \text{encode}(a_1, \ldots, a_n))\}$
    \item $IR_{variable} = \{(type: "variable", name: v)\}$
    \item $IR_{method} = \{(type: "method", obj: o, method: m, args: \text{encode}(a_1, \ldots, a_n))\}$
    \item $IR_{instantiate} = \{(type: "instantiate", class: c, args: \text{encode}(a_1, \ldots, a_n))\}$
    \item $IR_{delete} = \{(type: "delete", obj: o)\}$
\end{itemize}

\subsubsection{Encoding and Decoding Functions}

The encoding function $\text{encode}: \mathcal{V}_{L_s} \rightarrow \Sigma^*$ maps values from source language $L_s$ to JSON strings:

\begin{align}
\text{encode}(v) &= \begin{cases}
\text{JSON}(v) & \text{if } v \text{ is primitive} \\
\text{JSON}(\{varname: id, lang: L_s, typeName: \text{type}(v)\}) & \text{if } v \text{ is object}
\end{cases}
\end{align}

The decoding function $\text{decode}: \Sigma^* \rightarrow \mathcal{V}_{L_t}$ maps JSON strings to values in target language $L_t$:

\begin{align}
\text{decode}(s) &= \begin{cases}
\text{parse}(s) & \text{if } s \text{ is primitive} \\
\text{create\_reference}(s) & \text{if } s \text{ is object reference}
\end{cases}
\end{align}

\subsubsection{Global Variable Store}

The global variable store $\mathcal{G}_L$ for language $L$ maintains object references:

\begin{align}
\mathcal{G}_L &= \{id \mapsto (obj, type) \mid id \in \Sigma^*, obj \in \mathcal{O}_L, type \in \mathcal{T}_L\}
\end{align}

\subsubsection{Source-Level Transformation}

Given a source program $P_s$ in language $L_s$ containing foreign calls to language $L_t$, the transformation function $\mathcal{T}: P_s \rightarrow P_s'$ is defined as:

\begin{align}
\mathcal{T}(P_s) &= \text{rewrite}(P_s, \{\text{foreign\_call} \mapsto \text{runIR}(\text{generateIR}(\text{foreign\_call}))\})
\end{align}

where $\text{generateIR}$ maps foreign calls to IRs:

\begin{align}
\text{generateIR}(f_t(a_1, \ldots, a_n)) &= IR_{function}(f_t, \text{encode}(a_1, \ldots, a_n)) \\
\text{generateIR}(v_t) &= IR_{variable}(v_t) \\
\text{generateIR}(o_t.m(a_1, \ldots, a_n)) &= IR_{method}(o_t, m, \text{encode}(a_1, \ldots, a_n)) \\
\text{generateIR}(\text{new } C_t(a_1, \ldots, a_n)) &= IR_{instantiate}(C_t, \text{encode}(a_1, \ldots, a_n))
\end{align}

\subsubsection{IR Execution Semantics}

The execution of an IR in target kernel $K_t$ is defined by the function $\mathcal{E}: IR \times \mathcal{G}_{L_t} \rightarrow \mathcal{V}_{L_t} \times \mathcal{G}_{L_t}$:

\begin{align}
\mathcal{E}(IR_{function}(f, args), \mathcal{G}) &= (f(\text{decode}(args)), \mathcal{G}) \\
\mathcal{E}(IR_{variable}(v), \mathcal{G}) &= (\mathcal{G}[v], \mathcal{G}) \\
\mathcal{E}(IR_{method}(obj, m, args), \mathcal{G}) &= (\mathcal{G}[obj].m(\text{decode}(args)), \mathcal{G}) \\
\mathcal{E}(IR_{instantiate}(C, args), \mathcal{G}) &= (id, \mathcal{G} \cup \{id \mapsto (\text{new } C(\text{decode}(args)), C)\}) \\
\mathcal{E}(IR_{delete}(obj), \mathcal{G}) &= (\text{null}, \mathcal{G} \setminus \{obj\})
\end{align}

\subsubsection{Cross-Kernel Communication}

The cross-kernel communication function $\mathcal{C}: IR \times K_s \times K_t \rightarrow \mathcal{V}_{L_s}$ is defined as:

\begin{align}
\mathcal{C}(ir, K_s, K_t) &= \text{decode}(\text{send}(ir, K_t) \rightarrow \text{execute}(ir, K_t) \rightarrow \text{receive}(K_s))
\end{align}

where:
\begin{itemize}
    \item $\text{send}: IR \times K_t \rightarrow \text{Message}$ transmits IR to target kernel
    \item $\text{execute}: IR \times K_t \rightarrow \mathcal{V}_{L_t}$ executes IR in target kernel
    \item $\text{receive}: K_s \rightarrow \Sigma^*$ receives encoded result from target kernel
\end{itemize}

\subsubsection{Object Reference Management}

For object references, we define the reference creation function $\mathcal{R}: \mathcal{O}_{L_t} \times \mathcal{T}_{L_t} \rightarrow \text{Reference}$:

\begin{align}
\mathcal{R}(obj, type) &= \{varname: \text{generate\_id}(), lang: L_t, typeName: type\}
\end{align}

The reference resolution function $\mathcal{R}^{-1}: \text{Reference} \times \mathcal{G}_{L_t} \rightarrow \mathcal{O}_{L_t}$:

\begin{align}
\mathcal{R}^{-1}(ref, \mathcal{G}) &= \mathcal{G}[ref.varname]
\end{align}

\subsubsection{Semantic Preservation}

The transformation preserves semantics if for any source program $P_s$ and input $I$:

\begin{align}
\llbracket P_s \rrbracket(I) = \llbracket \mathcal{T}(P_s) \rrbracket(I)
\end{align}

where $\llbracket \cdot \rrbracket$ denotes the denotational semantics of the program.

\subsubsection{Type Safety}

Type safety is maintained if for any foreign call $f_t(a_1, \ldots, a_n)$ where $f_t: T_1 \times \ldots \times T_n \rightarrow R$:

\begin{align}
\text{type}(\text{decode}(\text{encode}(a_i))) &= T_i \quad \forall i \in [1, n] \\
\text{type}(\mathcal{C}(\text{generateIR}(f_t(a_1, \ldots, a_n)), K_s, K_t)) &= R
\end{align}

\subsubsection{State Consistency}

State consistency is maintained if for any sequence of operations $op_1, \ldots, op_n$:

\begin{align}
\mathcal{G}_{L_t}^n = \mathcal{E}(op_n, \mathcal{E}(op_{n-1}, \ldots, \mathcal{E}(op_1, \mathcal{G}_{L_t}^0) \ldots))
\end{align}

where $\mathcal{G}_{L_t}^i$ represents the global state after operation $i$.

This formal theory provides a rigorous foundation for understanding the semantics and correctness of the Kernel-FFI system, ensuring that cross-language function calls preserve meaning while maintaining type safety and state consistency across kernels.

\section{Implementation and Evaluation}
\label{sec:evaluation}

In this section, we evaluate the proposed approach by implementing the framework and supporting different programming languages. We describe the implementation details for various programming languages and highlight how different language features are supported.

\subsection{Unified Framework supporting multiple languages and kernels}

\begin{figure}
  \centering
  \includegraphics[width=0.8\linewidth]{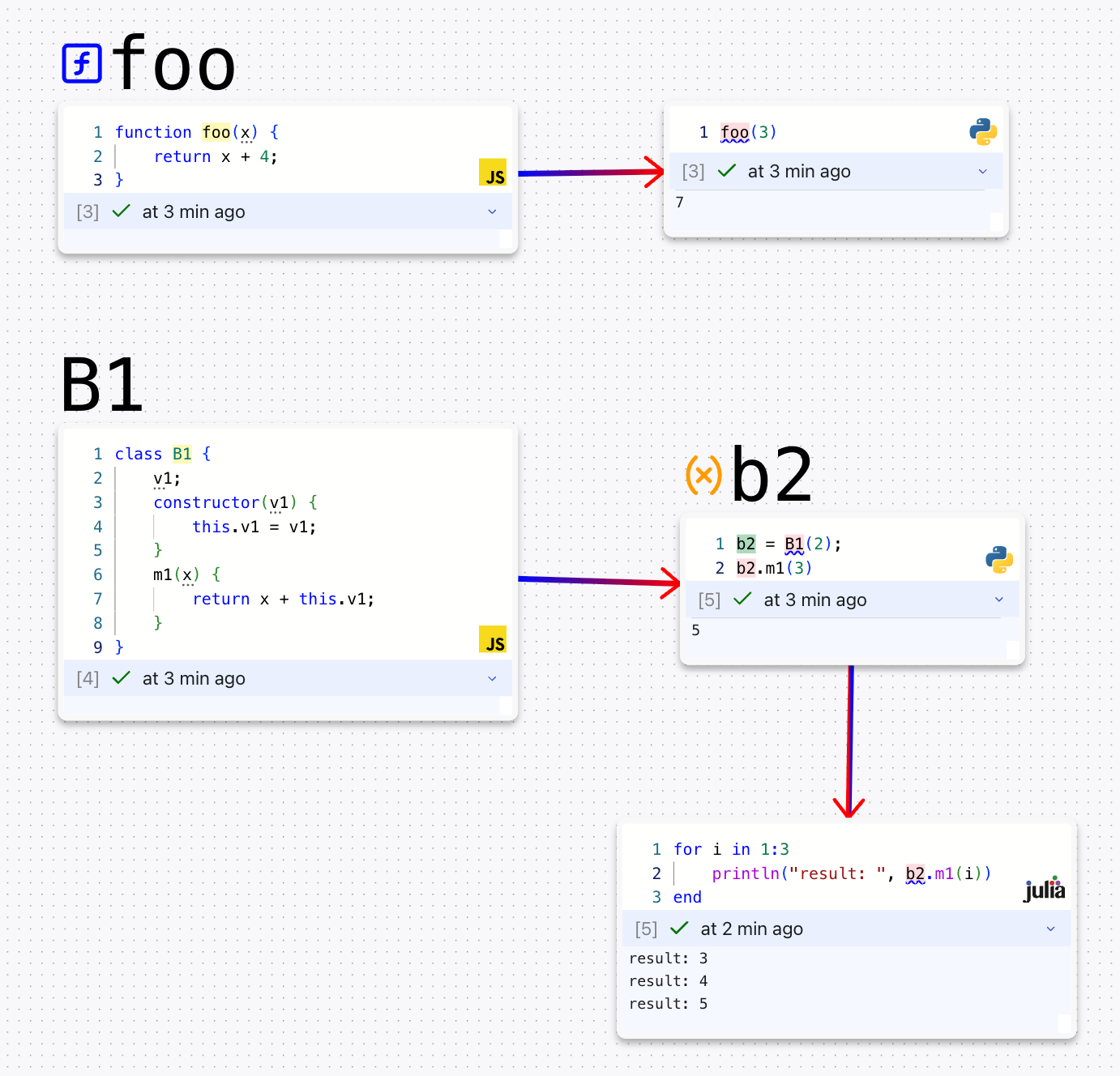}
  \caption{Unified framework in Kernel-FFI enables interactive development of code in multiple languages and kernels, reducing the effort required to manage cross-language interactions.}
  \label{fig:codepod-unified-framework}
\end{figure}

To better develop code in different languages simultaneously, we introduce a unified framework that allows developers to interactively develop code in multiple languages and kernels. The framework is similar to Jupyter, except that developers can (1) create code blocks on a 2D canvas instead of a linear notebook, and (2) create code cells in different languages. A screenshot of the unified framework is shown in Figure~\ref{fig:codepod-unified-framework}.

The unified framework automatically parses each code cell of each language and registers defined top-level functions, classes, and variables into the symbol table registry. If a symbol is defined in one language and used in another language, we automatically create the foreign function call and generate IRs and target code for executing the foreign calls. This feature significantly reduces development effort and improves productivity of multi-language projects.

\subsection{Programming Languages of Various Types}

\begin{table}[ht]
  \footnotesize
  \centering
  \begin{tabular}{l|c|c|c|l|l}
    \toprule

  Language & \begin{tabular}[c]{@{}c@{}}Compile-Time\\Type Checking\end{tabular} & Eval & Vararg & Kernel & Features \\
  \midrule
  Python & \cross & \tick & \tick & ipython~\cite{ipython_kernel} & data science, ML/AI, web dev, automation \\
  JS/TS & \cross & \tick & \tick & deno~\cite{deno_kernel} & web frontend/backend, Node.js, async/await \\
  Julia & \cross & \tick & \tick & IJulia~\cite{ijulia_kernel} & scientific computing, multiple dispatch, macros \\
  Racket & \cross & \tick & \tick & iracket~\cite{iracket_kernel} & functional prog, hygienic macros, DSLs \\
  Ruby & \cross & \tick & \tick & iruby~\cite{iruby_kernel} & web dev (Rails), meta-prog, elegant syntax \\
  C\# & \tick & \cross & \tick & dotnet-inter~\cite{dotnet_interactive_kernel} & .NET ecosystem, Windows dev, game dev (Unity) \\
  C++ & \tick & \cross & \cross & xeus-cling~\cite{xeus_cling_kernel} & systems prog, perf-critical apps, templates \\
  Go & \tick & \cross & \cross & gophernotes~\cite{gophernotes_kernel} & cloud services, microservices, goroutines, simplicity \\
  Rust & \tick & \cross & \cross & evcxr\_jupyter~\cite{evcxr_jupyter_kernel} & memory safety, ownership, systems prog, zero-cost abs \\
  \bottomrule
  \end{tabular}
  \caption{Programming languages of various types implemented for Kernel-FFI.}
  \label{tab:programming-languages}
\end{table}

Table~\ref{tab:programming-languages} shows the programming languages currently implemented to support Kernel-FFI, along with their key characteristics and features that influence the implementation. We categorize languages based on several important properties: compile-time type checking, eval support, vararg support, kernel implementation, and distinctive features. Languages like C++, Rust, Go, and C\# perform type checking at compile time and require explicit type annotations, while dynamic languages like Python, JavaScript, Julia, Racket, and Ruby support runtime evaluation and more flexible typing. This affects how Kernel-FFI handles type information and implements foreign function calls. These language features are discussed in detail in the following sections when we introduce the differences in supporting Kernel-FFI for different languages.

The diverse set of languages demonstrates Kernel-FFI's ability to support both static and dynamic typing, different execution models, and various programming paradigms—from Python's data science ecosystem to Rust's memory safety guarantees. The implementation adapts to each language's unique characteristics while maintaining consistent cross-language interaction semantics.

\subsection{Host and Client Code for Different Languages}

\begin{figure}[ht]
  \centering
  \begin{subfigure}[b]{0.19\linewidth}
    \centering
    \includegraphics[width=\linewidth]{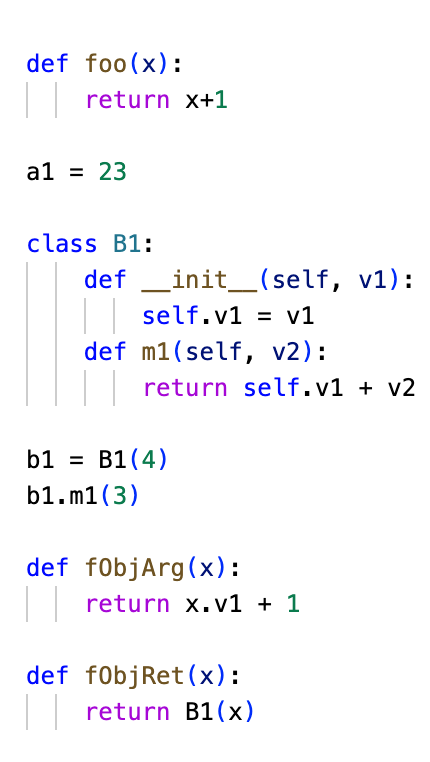}
    \caption{Python}
    \label{fig:python-host}
  \end{subfigure}
  \hfill
  \begin{subfigure}[b]{0.19\linewidth}
    \centering
    \includegraphics[width=\linewidth]{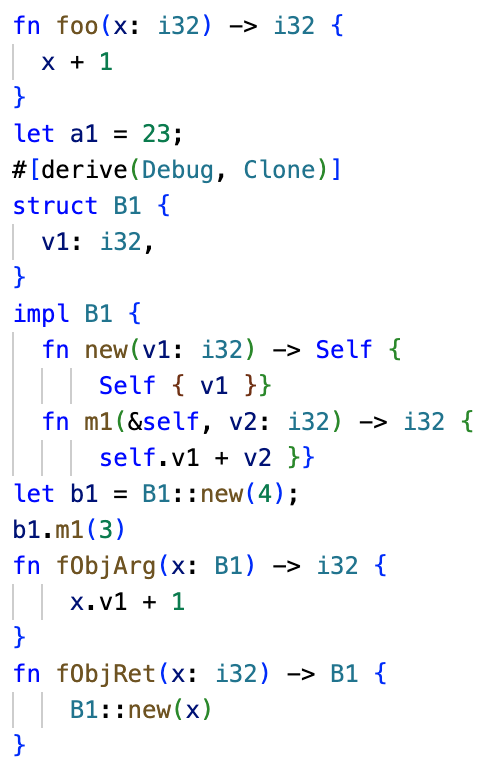}
    \caption{Rust}
    \label{fig:rust-host}
  \end{subfigure}
  \hfill
  \begin{subfigure}[b]{0.19\linewidth}
    \centering
    \includegraphics[width=\linewidth]{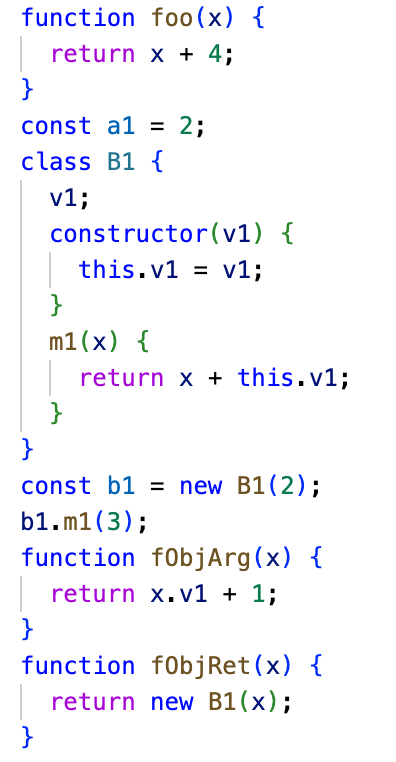}
    \caption{TypeScript}
    \label{fig:ts-host}
  \end{subfigure}
  \hfill
  \begin{subfigure}[b]{0.19\linewidth}
    \centering
    \includegraphics[width=\linewidth]{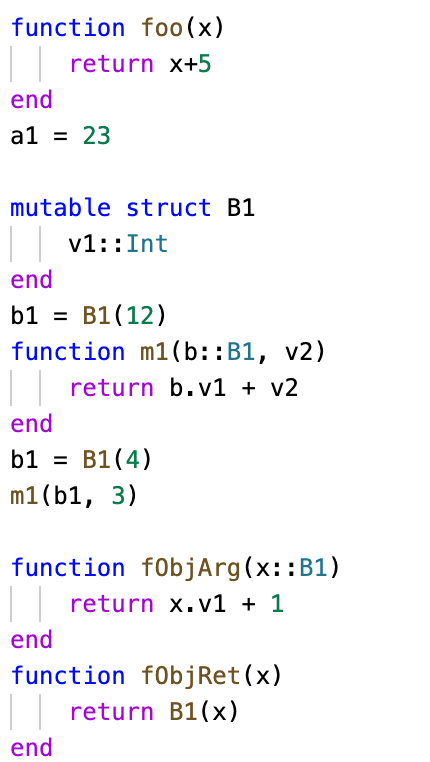}
    \caption{Julia}
    \label{fig:julia-host}
  \end{subfigure}
  \hfill
  \begin{subfigure}[b]{0.19\linewidth}
    \centering
    \includegraphics[width=\linewidth]{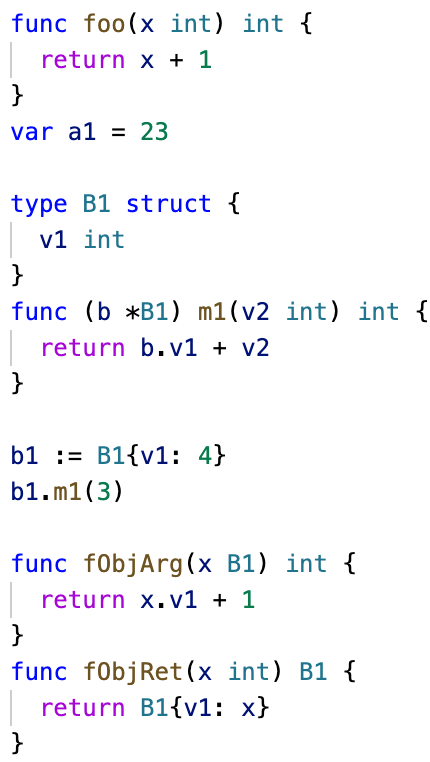}
    \caption{Go}
    \label{fig:go-host}
  \end{subfigure}
  \hfill
  \begin{subfigure}[b]{0.19\linewidth}
    \centering
    \includegraphics[width=\linewidth]{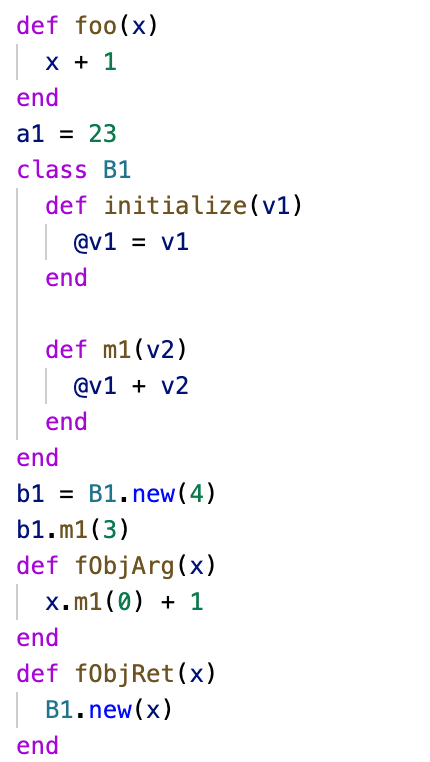}
    \caption{Ruby}
    \label{fig:ruby-host}
  \end{subfigure}
  \hfill
  \begin{subfigure}[b]{0.19\linewidth}
    \centering
    \includegraphics[width=\linewidth]{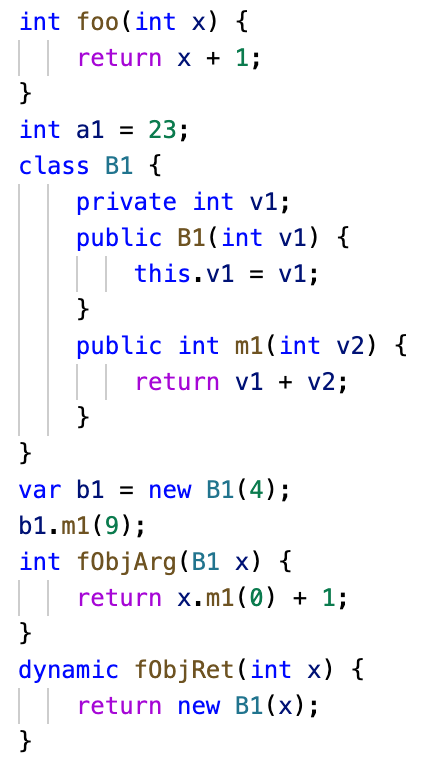}
    \caption{C\#}
    \label{fig:csharp-host}
  \end{subfigure}
  \hfill
  \begin{subfigure}[b]{0.19\linewidth}
    \centering
    \includegraphics[width=\linewidth]{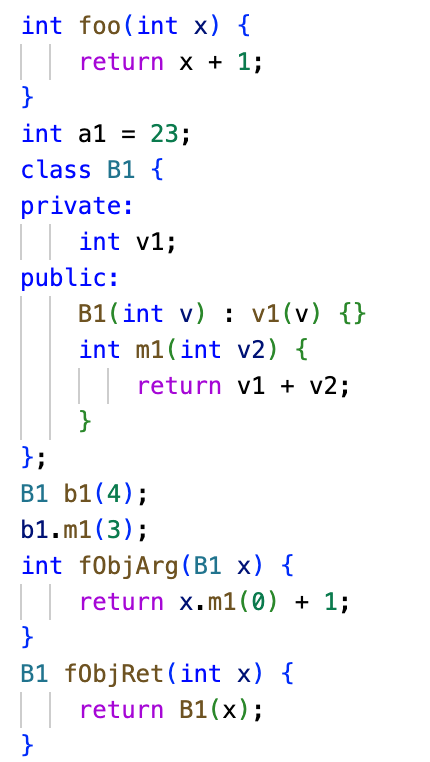}
    \caption{C++}
    \label{fig:cpp-host}
  \end{subfigure}
  \hfill
  \begin{subfigure}[b]{0.19\linewidth}
    \centering
    \includegraphics[width=\linewidth]{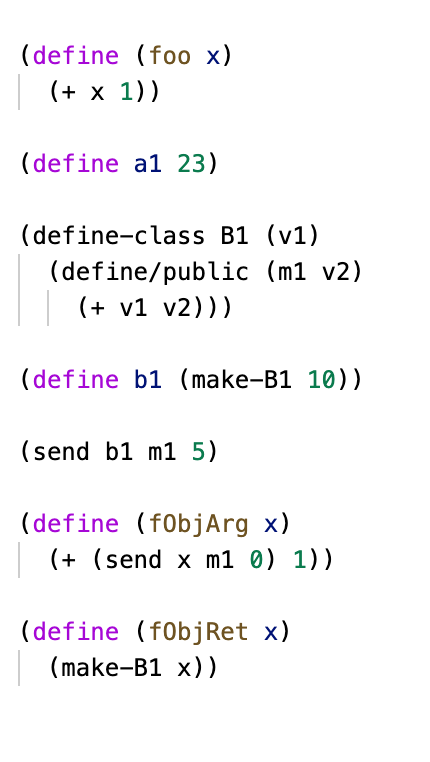}
    \caption{Racket}
    \label{fig:racket-host}
  \end{subfigure}
  \hfill
  \begin{subfigure}[b]{0.19\linewidth}
    \centering
    \phantom{\includegraphics[width=\linewidth]{assets-lang/host-racket.png}}
  \end{subfigure}
  \caption{Example of host code that defines functions, variables, and classes for different languages.}
  \label{fig:host-code}
\end{figure}

\begin{figure}[ht]
    \centering
    \begin{subfigure}[b]{0.19\linewidth}
      \centering
      \includegraphics[width=\linewidth]{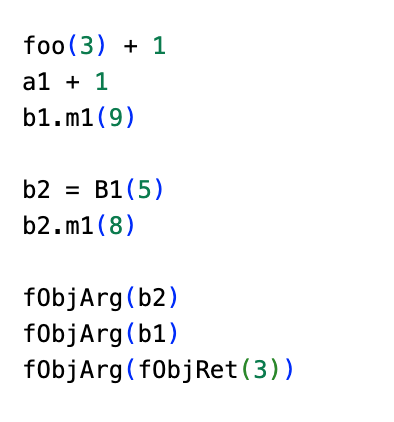}
      \caption{Python}
      \label{fig:python-client}
    \end{subfigure}
    \hfill
    \begin{subfigure}[b]{0.19\linewidth}
      \centering
      \includegraphics[width=\linewidth]{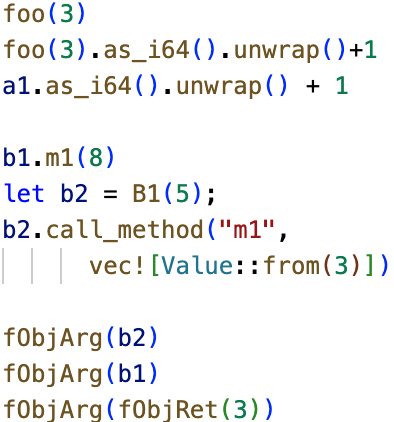}
      \caption{Rust}
      \label{fig:rust-client}
    \end{subfigure}
    \hfill
    \begin{subfigure}[b]{0.19\linewidth}
      \centering
      \includegraphics[width=\linewidth]{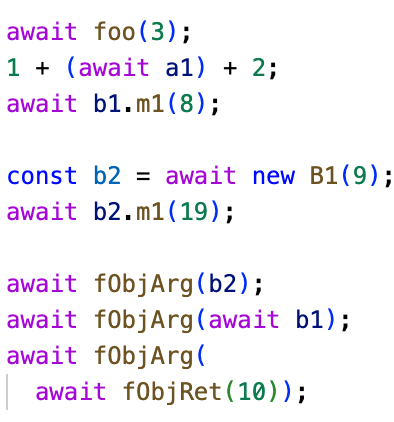}
      \caption{TypeScript}
      \label{fig:ts-client}
    \end{subfigure}
    \hfill
    \begin{subfigure}[b]{0.19\linewidth}
      \centering
      \includegraphics[width=\linewidth]{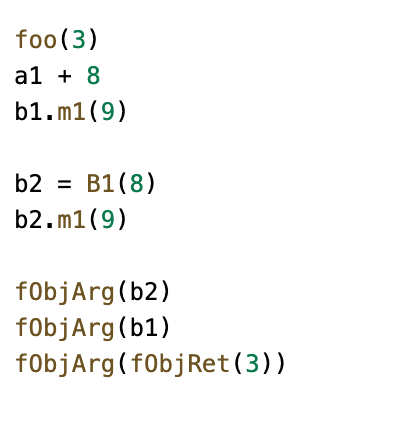}
      \caption{Julia}
      \label{fig:julia-client}
    \end{subfigure}
    \hfill
    \begin{subfigure}[b]{0.19\linewidth}
      \centering
      \includegraphics[width=\linewidth]{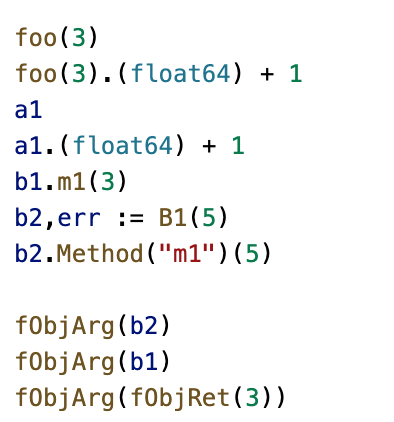}
      \caption{Go}
      \label{fig:go-client}
    \end{subfigure}
    \hfill
    \begin{subfigure}[b]{0.19\linewidth}
      \centering
      \includegraphics[width=\linewidth]{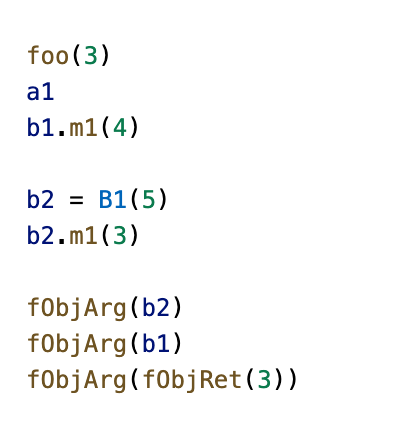}
      \caption{Ruby}
      \label{fig:ruby-client}
    \end{subfigure}
    \hfill
    \begin{subfigure}[b]{0.19\linewidth}
      \centering
      \includegraphics[width=\linewidth]{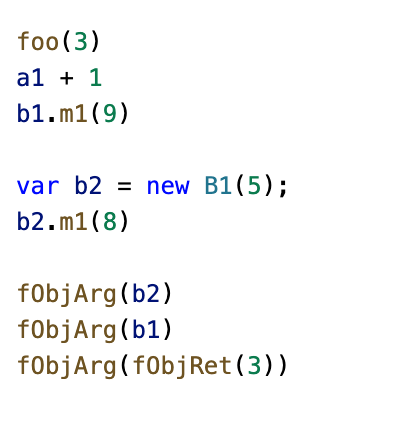}
      \caption{C\#}
      \label{fig:csharp-client}
    \end{subfigure}
    \hfill
    \begin{subfigure}[b]{0.19\linewidth}
      \centering
      \includegraphics[width=\linewidth]{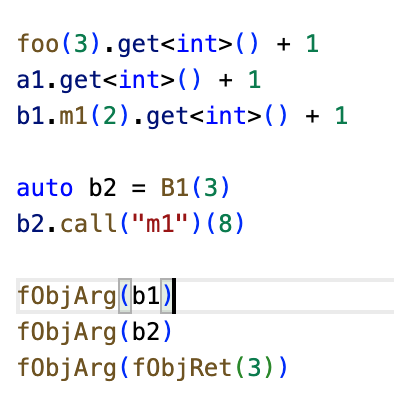}
      \caption{C++}
      \label{fig:cpp-client}
    \end{subfigure}
    \hfill
    \begin{subfigure}[b]{0.19\linewidth}
      \centering
      \includegraphics[width=\linewidth]{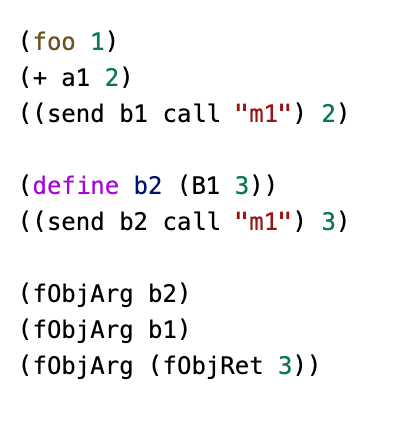}
      \caption{Racket}
      \label{fig:racket-client}
    \end{subfigure}
    \hfill
    \begin{subfigure}[b]{0.19\linewidth}
      \centering
      \phantom{\includegraphics[width=\linewidth]{assets-lang/client-racket.png}}
    \end{subfigure}
    \caption{Example of client code that calls remote functions, variables, and classes defined in the host code for different languages. The code is further analyzed in Section \ref{sec:further-client-computation} for how to use the return values in further client computation.}
    \label{fig:client-code}
\end{figure}

Figure~\ref{fig:host-code} shows examples of host programs that define functions, variables, and classes for different languages. The host code is written in the host language, and is executed in the host kernel. The host programs of different languages are semantically equivalent. Figure~\ref{fig:client-code} shows examples of client programs that call remote functions, variables, and classes defined in the host code for different languages. The client code is written in the client language, and is executed in the client kernel. The client programs of different languages are semantically equivalent.

\subsection{Eval endpoints: Jupyter wire and Language eval}
\label{sec:eval-endpoint}

The rewritten code is executed in the Jupyter kernel. We use the side-channel to communicate with the kernel to enable non-blocking recursive FFI executions. This feature requires the language runtime to support evaluation of arbitrary code. This is supported by dynamic languages, including Python, JS/TS, Ruby, Julia, and Racket.

\begin{figure}[ht]
    \centering
    \begin{subfigure}[b]{0.45\linewidth}
      \centering
      \includegraphics[width=\linewidth]{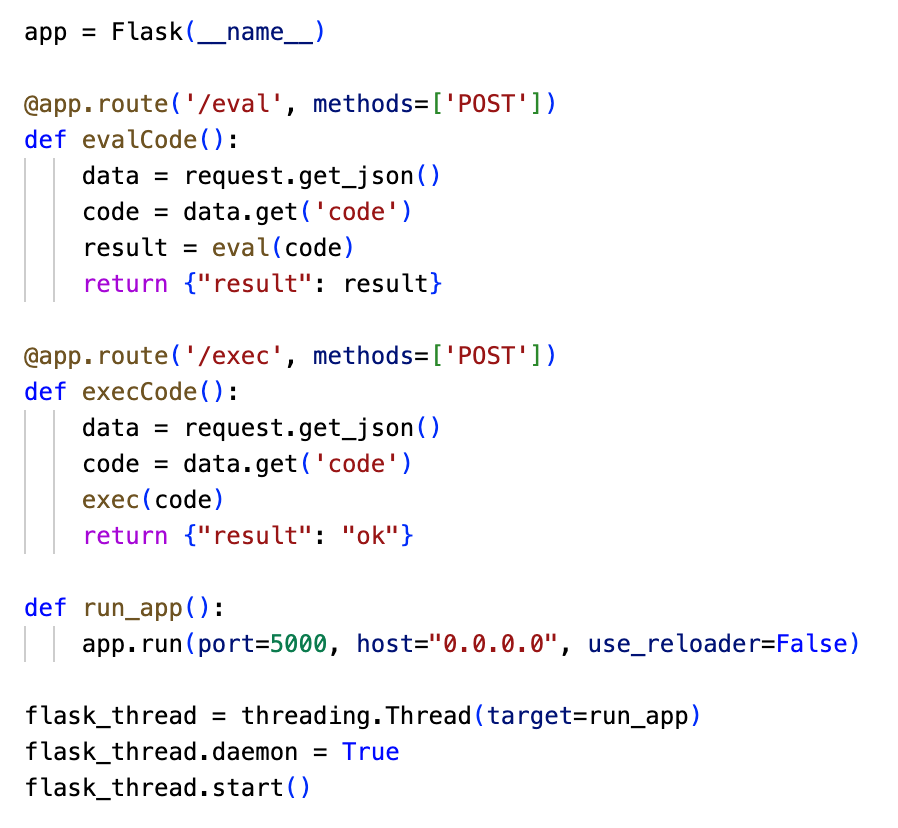}
      \caption{Python}
      \label{fig:python-eval}
    \end{subfigure}
    \hfill
    \begin{subfigure}[b]{0.45\linewidth}
      \centering
      \includegraphics[width=\linewidth]{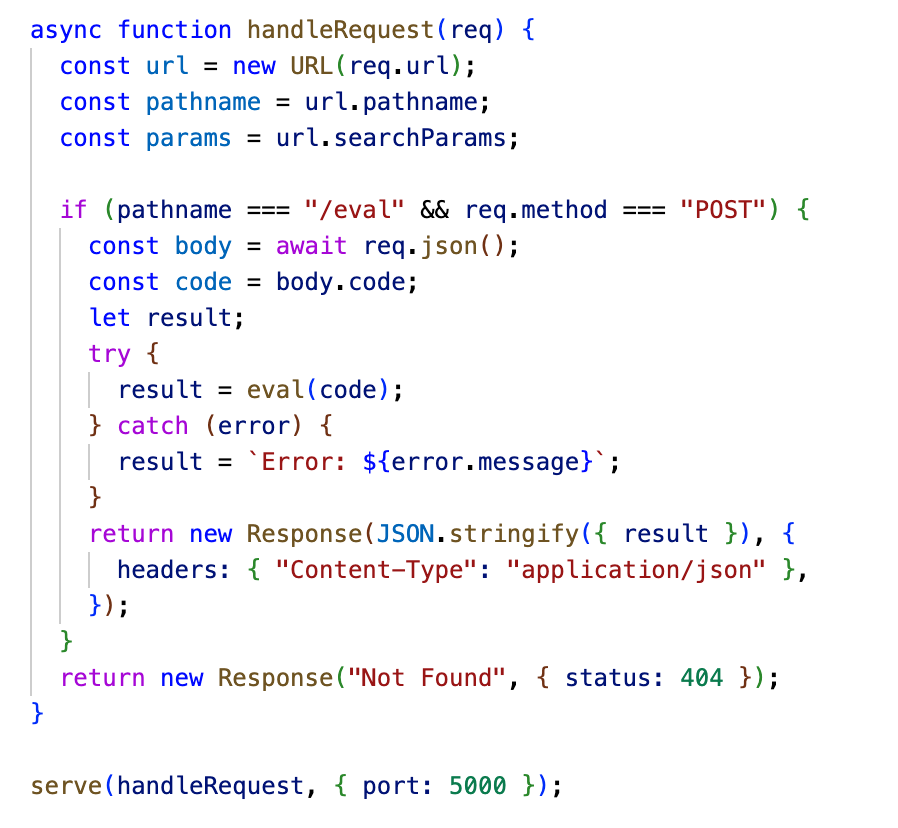}
      \caption{Typescript}
      \label{fig:ts-eval}
    \end{subfigure}

    \caption{Example of eval code for python and typescript.}
    \label{fig:eval-code}
  \end{figure}

Python has two endpoints due to the fact that Python has \texttt{eval} and \texttt{exec}, with differences. \texttt{eval} returns a value, while \texttt{exec} runs a statement for effects. \texttt{exec} is used when the IR type is instantiate and delete, and \texttt{eval} is used for all other IR types. For other languages (JS/TS, Julia, Ruby, Racket), just one eval endpoint is used for all IR types. The eval endpoints for Python and TypeScript are shown in Figure~\ref{fig:eval-code}.

For languages that do not support evaluation of arbitrary code (C\#, Go, C++, Rust), we fall back to the Jupyter wire protocol. The Jupyter wire protocol is a WebSocket connection between the client and the kernel. The client sends the rewritten code to the kernel, and the kernel executes the code and sends the result back to the client.

\subsection{Type casting in args decoding and globalVars retrieval}

When generating the target code from IR, we need to first decode the arguments, and then call the function on the decoded arguments. Calling functions with arbitrary numbers and types of arguments differs across languages. This is related to whether the language supports variable arguments and whether the language requires static typing at compile time instead of dynamic typing at runtime.

\subsubsection{Implicit dynamic type casting}
\begin{figure}[ht]
    \centering
    \begin{subfigure}[b]{0.45\linewidth}
      \centering
      \includegraphics[width=\linewidth]{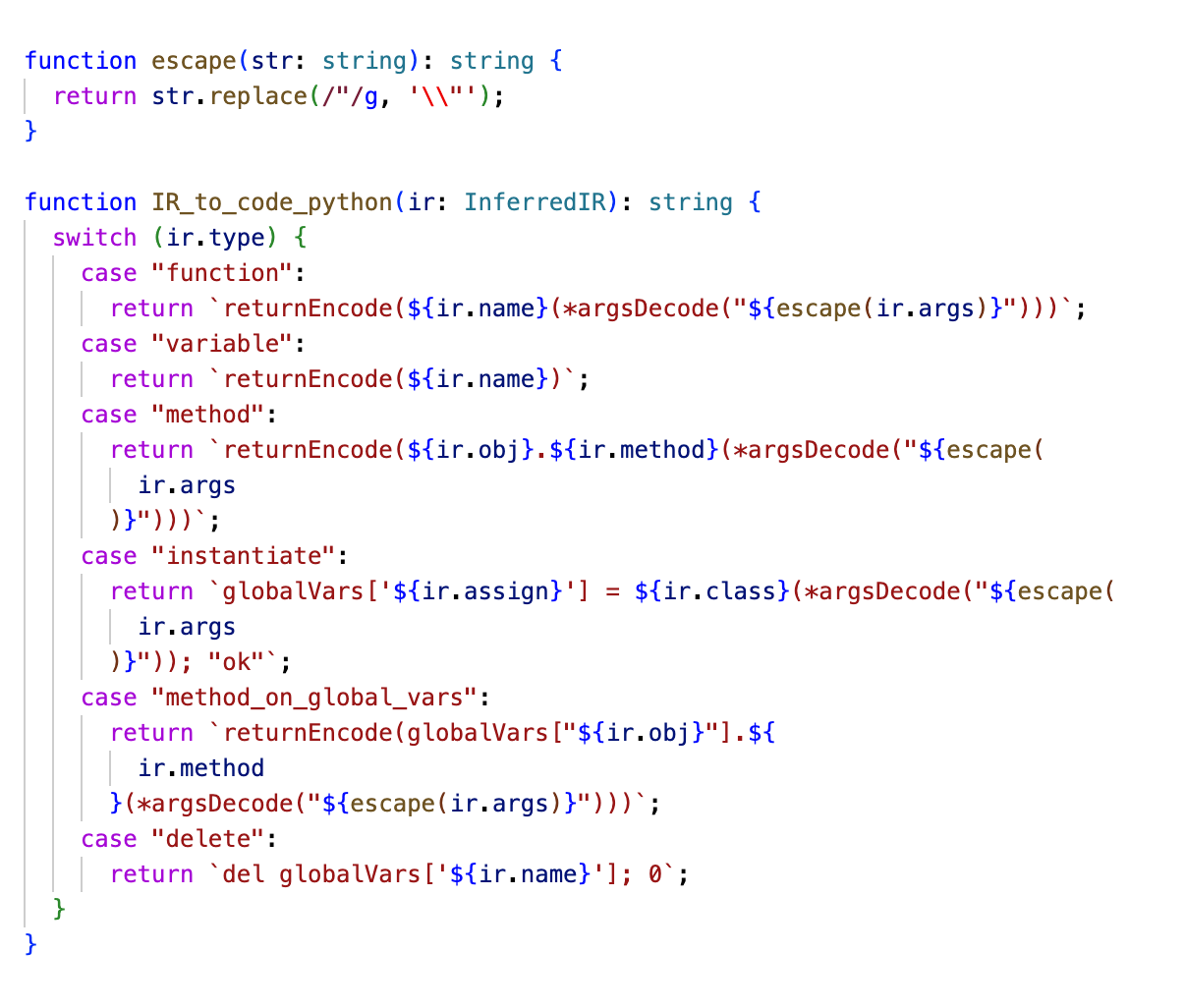}
      \caption{IR to Python code generation in intermediate system (typescript)}
      \label{fig:python-IR_to_code}
    \end{subfigure}
    \hfill
    \begin{subfigure}[b]{0.45\linewidth}
      \centering
      \includegraphics[width=\linewidth]{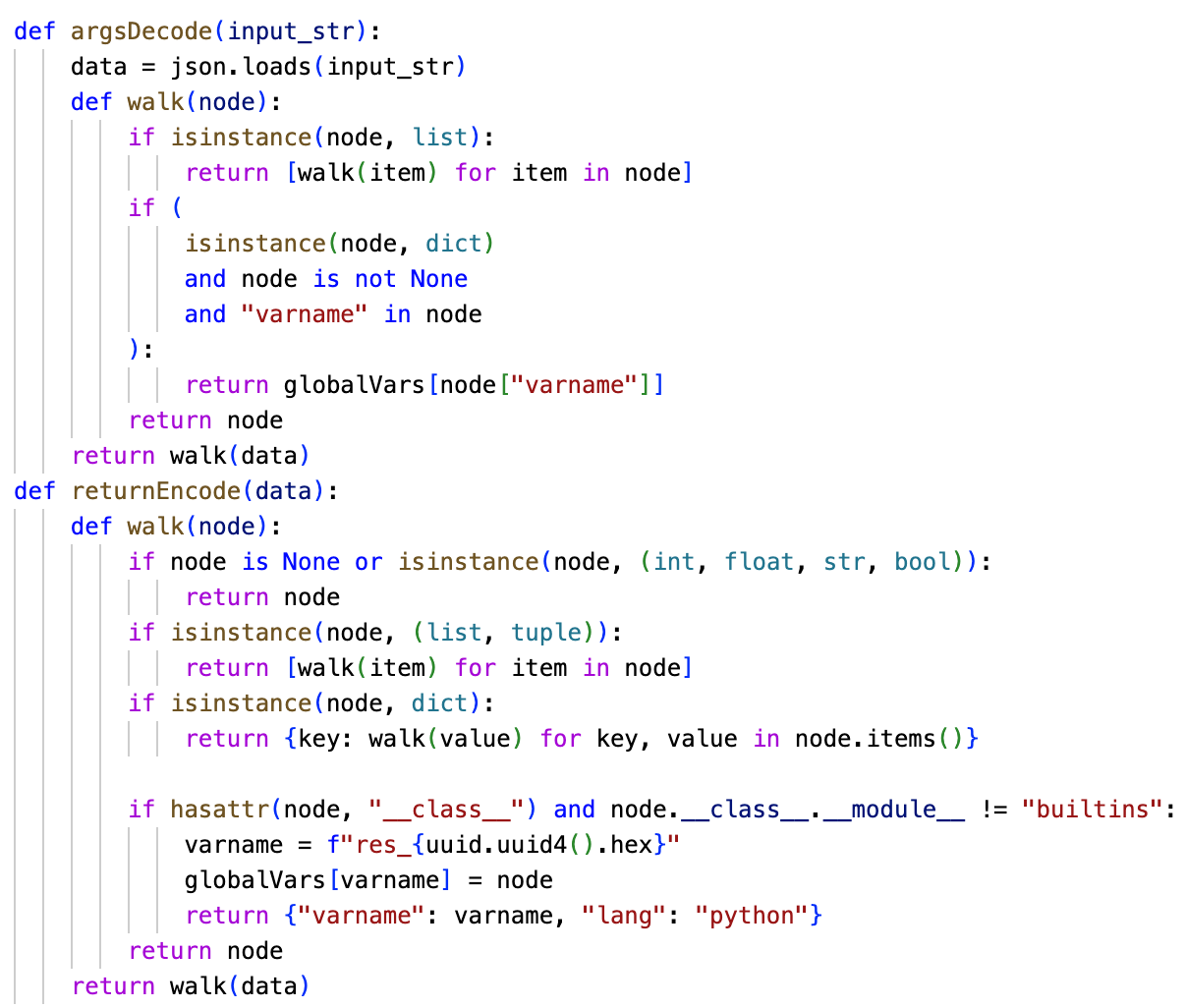}
      \caption{Python Host Args decoding and result encoding}
      \label{fig:python-argsDecode}
    \end{subfigure}

    \caption{Dynamic type checked language: IR to code generation and args decoding for python.}
    \label{fig:python-IR_to_code-argsDecode}
  \end{figure}

Dynamic languages including Python, JS/TS, Ruby, Julia, and Racket support variable arguments and also perform dynamic type checking at runtime. Therefore, we can use the \texttt{*args} syntax to pass variable arguments to a function. Also, the argsDecoding is implemented in the host language to natively parse the JSON string in the IR args field and convert it into objects in the host kernel runtime. The IR to code generation and args decoding for Python are shown in Figure~\ref{fig:python-IR_to_code-argsDecode}.

\subsubsection{Explicit static type casting}

For statically typed languages (C++, Go, Rust, C\#) and languages that do not support variable arguments (C++, Go, Rust), we need to decode the arguments in our system instead of inside the host language before sending to the kernel for execution. This is done when generating the target code from IR. Specifically, our intermediate system that transforms the IR to code is implemented as a TypeScript server, and we can use the TypeScript runtime to decode the arguments and generate code that calls the function on the decoded arguments.

\begin{figure}[ht]
    \centering
    \begin{subfigure}[b]{0.49\linewidth}
      \centering
      \includegraphics[width=\linewidth]{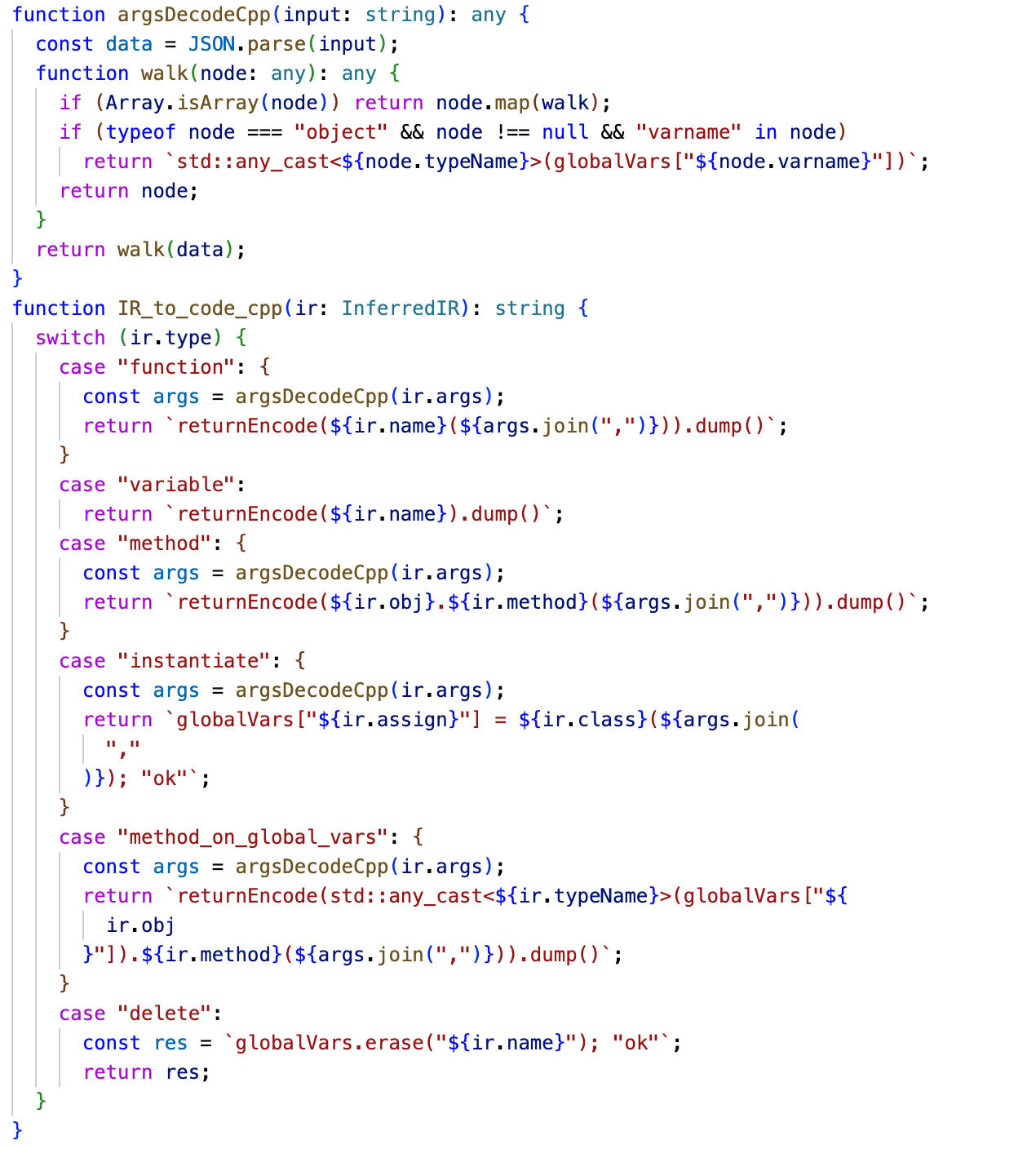}
      \caption{IR to C++ code generation in intermediate system (typescript)}
      \label{fig:cpp-IR_to_code}
    \end{subfigure}
    \hfill
    \begin{subfigure}[b]{0.49\linewidth}
      \centering
      \includegraphics[width=\linewidth]{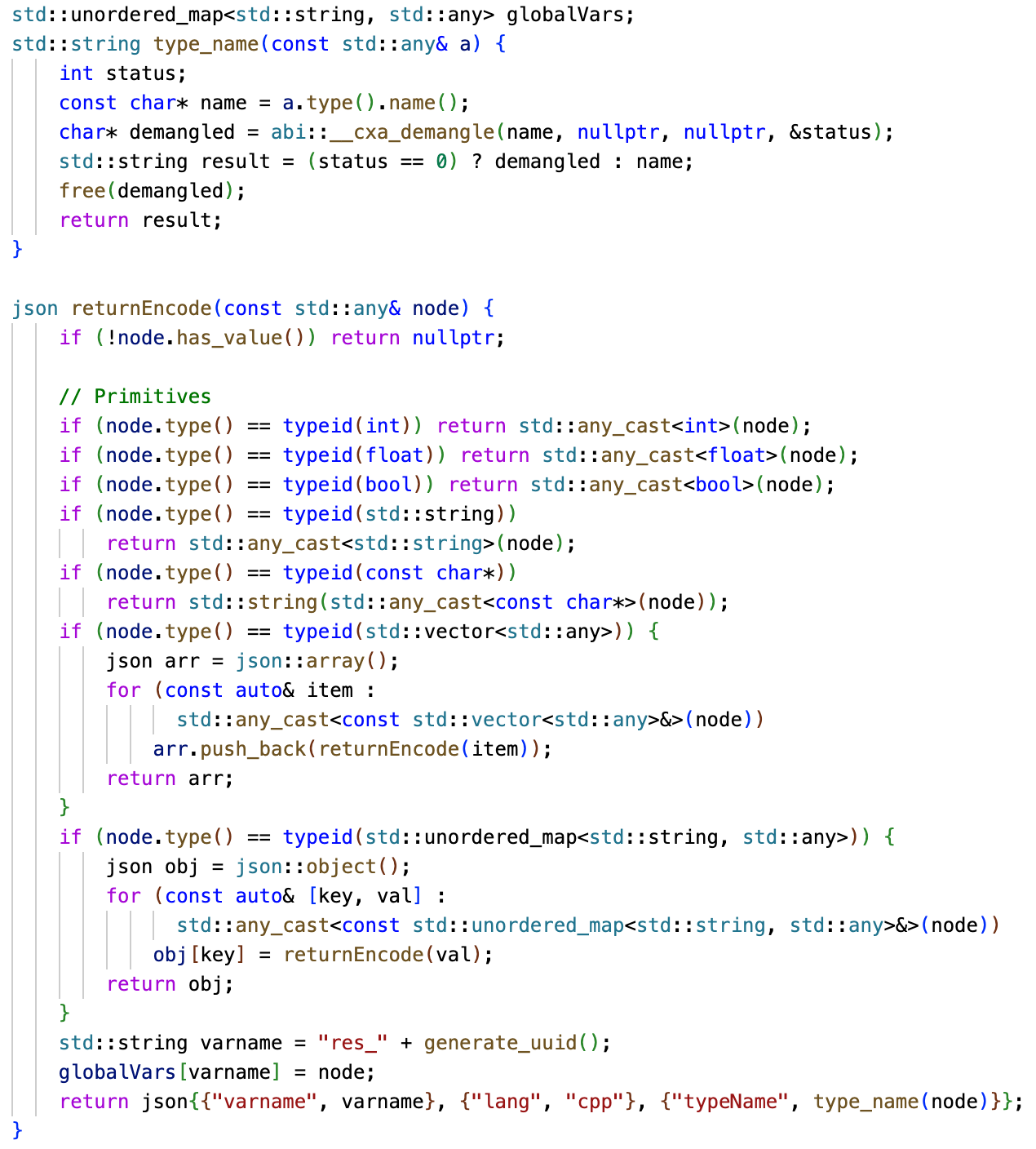}
      \caption{Return encoding in C++ host}
      \label{fig:cpp-returnEncode}
    \end{subfigure}

    \caption{Statically typed language: IR to code generation and globalVars casting for C++.}
    \label{fig:cpp-IR_to_code-returnEncode}
  \end{figure}

In these statically typed languages, we also need to cast the globalVars to the target type when it is retrieved by the remote object referencing in the client code. To get the type information, we use introspection or reflection provided by the language runtime when encoding the return results from the host kernel. For example, the IR to code generation and globalVars casting for C++ are shown in Figure~\ref{fig:cpp-IR_to_code-returnEncode}.

\subsubsection{Explicit dynamic type casting}

\begin{figure}[ht]
  \centering
  \begin{subfigure}[b]{0.49\linewidth}
    \centering
    \includegraphics[width=\linewidth]{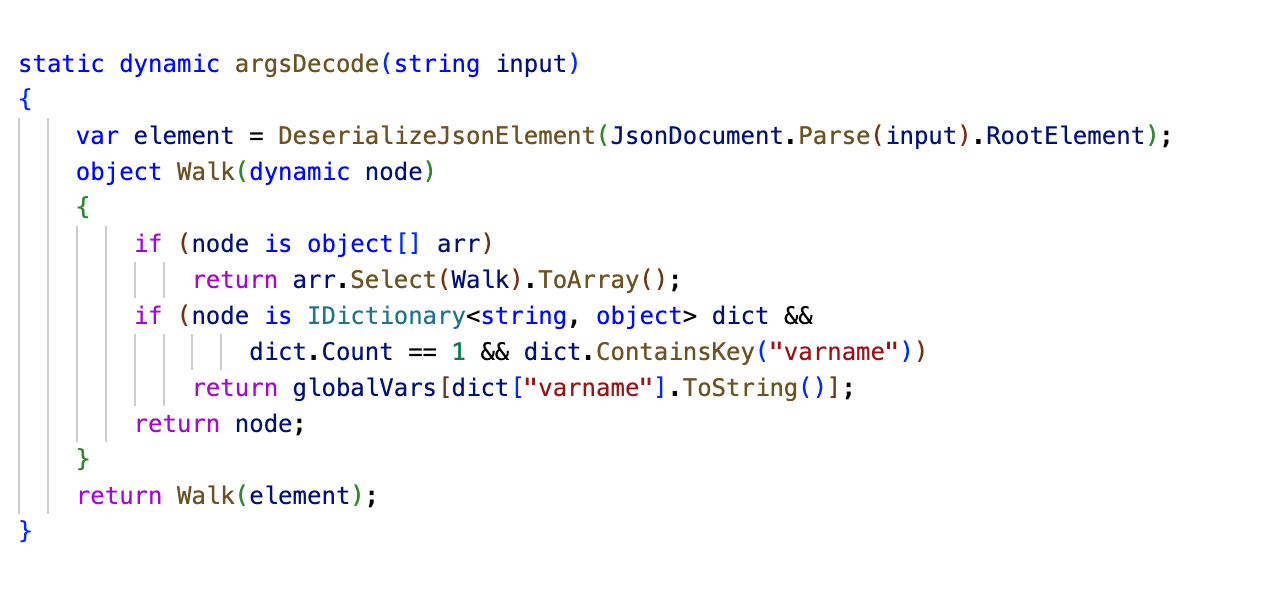}
    \caption{C\# Host Args decoding and result encoding}
    \label{fig:csharp-argsDecode}
  \end{subfigure}
  \hfill
  \begin{subfigure}[b]{0.49\linewidth}
    \centering
    \includegraphics[width=\linewidth]{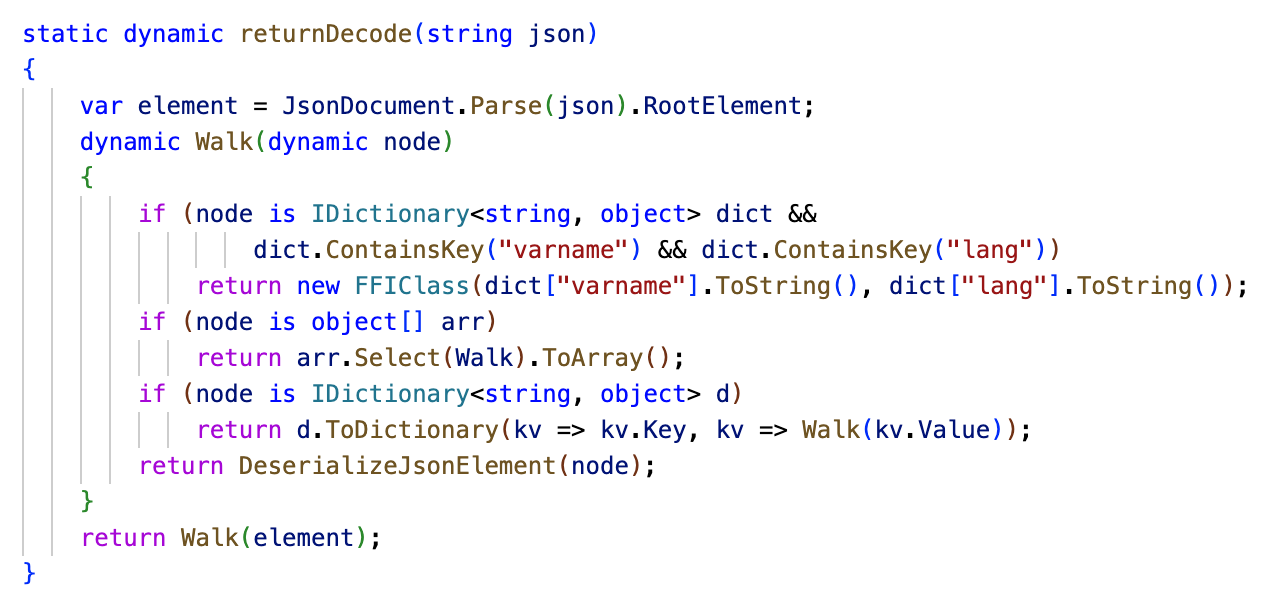}
    \caption{Return decoding in C\# client}
    \label{fig:csharp-returnDecode}
  \end{subfigure}

  \caption{Explicit dynamic type casting for C\# in argsDecode and returnDecode.}
  \label{fig:csharp-explicit-dynamic-casting}
\end{figure}

For C\#, the language supports the \texttt{dynamic} keyword, which allows runtime type checking. We use this keyword to indicate the type of the arguments and return values. The argsDecode function for C\# is shown in Figure~\ref{fig:csharp-explicit-dynamic-casting}.

\subsection{Return encoding and use in further client computation}
\label{sec:further-client-computation}

\paragraph{Implicit dynamic type casting in returnDecode}
The dynamic languages including Python, JS/TS, Ruby, Julia, and Racket have type checking at runtime, so the return values, once decoded by returnDecode, are native objects in the client kernel that are ready to use in further client computation.

\paragraph{Explicit dynamic type casting in returnDecode} In C\#, although the language is statically typed, the \texttt{dynamic} keyword is used to indicate the type of the arguments and return values. This allows runtime type checking. Therefore, we can embed the explicit type casting in the returnDecode function and the client code will not need to do any type casting and the result will be a native object in the client kernel directly usable for further computation. The returnDecode function for C\# is shown in Figure~\ref{fig:csharp-explicit-dynamic-casting}.

\paragraph{Explicit static type casting in returnDecode}
Rust, Go, and C++ are statically typed languages, so the return values are encoded as JSON strings in the IR. To use the return values in further client computation, we need to cast the JSON object to the target type. For example, in Rust in Figure~\ref{fig:rust-client}, the return value of \texttt{foo(3)} needs to be cast to the desired type by \texttt{foo(3).as\_i64().unwrap()} and then used in computation \texttt{foo(3).as\_i64().unwrap() + 1}.

\paragraph{Async return value in returnDecode}
For TypeScript and JavaScript, the remote FFI call is implemented as an async function call, and the return value is a promise. The client code will need to await the promise in Jupyter's top-level to resolve and then use the resolved value in further client computation. For example, the return value of \texttt{foo(3)} in TypeScript in Figure~\ref{fig:ts-client} is a promise, and the client code will need to run \texttt{await foo(3)} to get the resolved value and then use it in further computation like \texttt{await foo(3) + 1}.

\section{Related Work}
\label{sec:related-work}

The problem of enabling interaction between programming languages has been a longstanding area of research and development. Various techniques, ranging from traditional Foreign Function Interfaces (FFIs) to Remote Procedure Calls (RPCs), have been employed to achieve cross-language interoperability. This section reviews these techniques and highlights their limitations, which Kernel-FFI aims to address.

\subsection{Foreign Function Interfaces (FFIs)}

Traditional FFIs enable one language to call functions or use libraries written in another language, typically C or C-compatible languages. Examples include Python's \texttt{ctypes} library, Java's Java Native Interface (JNI), and the Rust Foreign Function Interface. These systems rely on language bindings to define how data structures and function calls are translated between the source and target languages.

While FFIs are powerful, they are often restricted to pairs of languages and require significant manual effort. For instance, developers must create bindings for each function or library they wish to use, which involves understanding the memory layout, type systems, and calling conventions of both languages. This process is error-prone and can introduce subtle bugs related to memory management and type mismatches~\cite{ffi_problems}. Moreover, most FFIs are limited to static, language-level interoperability and do not support dynamic or multi-language workflows.

\subsection{Remote Procedure Calls (RPCs)}

RPC frameworks, such as gRPC~\cite{grpc}, Thrift~\cite{thrift}, and XML-RPC, provide an alternative approach to cross-language communication. These systems allow functions in one language to be invoked from another by transmitting serialized data over a network. RPC frameworks typically use Interface Definition Languages (IDLs) to define function signatures and data types, which are then compiled into stubs for each target language. gRPC are highly scalable and are widely used in distributed systems.

However, RPC frameworks require predefined schemas, making them less suitable for interactive or exploratory development environments, such as Jupyter notebooks. gRPC needs heavy boilerplate code which is not friendly for interactive Jupyter sessions. Our proposed method is transparent and does not require any boilerplate code. Different languages can directly call each other as if native calls. Second, gRPC only supports functions and primitive types. It doesn't support class instantiation, custom type objects, and states (top-level variables), which are commonly used and frequently updated in interactive development sessions in Jupyter. We support not only functions and primitive types, but also class instantiation, custom objects, states (remote variable referencing).


\subsection{REST Web APIs}

REST (Representational State Transfer) web APIs are another widely used approach for cross-language communication. REST APIs expose language-agnostic endpoints, typically using HTTP, to enable the exchange of data between systems. Languages such as Python, JavaScript, and Java can interact seamlessly by implementing HTTP requests and responses in accordance with REST conventions. Frameworks like Flask, Express, and Spring Boot simplify the process of creating and consuming RESTful services.

While REST APIs provide simplicity and broad compatibility, they introduce significant latency due to the stateless nature of HTTP and the overhead of request parsing and response generation. Furthermore, REST APIs are primarily designed for coarse-grained communication between services, making them less suitable for fine-grained function calls or workflows requiring frequent state synchronization. Unlike Kernel-FFI, REST APIs lack the ability to manage in-memory state across languages or provide the seamless interactivity needed in development environments like Jupyter.

\subsection{Cross-Language Virtual Machines}

Cross-language virtual machines, such as the Java Virtual Machine (JVM)~\cite{jvm} and Microsoft's Common Language Runtime (CLR)~\cite{clr}, provide a unified runtime for executing programs written in multiple languages. For example, Scala, Kotlin, and Groovy can all run on the JVM, and .NET supports languages such as C\#, F\#, and VB.NET. These systems achieve interoperability by compiling all supported languages into a common intermediate representation.

While effective for tightly coupled ecosystems, these approaches are inherently limited to languages designed to target the same runtime. They do not address the broader need for interoperability between languages with distinct runtimes, such as Python, JavaScript, and R. Kernel-FFI addresses this gap by enabling communication between languages without requiring a shared runtime.

\subsection{Polyglot Systems and Language Servers}

Recent advancements in polyglot systems, such as GraalVM~\cite{graalvm}, have attempted to unify languages under a single execution environment. GraalVM supports languages like JavaScript, Python, and R, allowing them to interoperate through a common interface. However, these systems often require modifications to the underlying languages or runtime environments, limiting their generality.

In the context of development environments, language servers such as Microsoft's Language Server Protocol (LSP)~\cite{lsp} provide a standardized way to interact with multiple programming languages. While LSP focuses on development features like autocompletion and linting, Kernel-FFI extends this concept to runtime interoperability, enabling seamless execution of code across different kernels.

\subsection{Multi-Language Toolkits}

Several multi-language toolkits, such as Apache Arrow~\cite{arrow}, aim to facilitate data sharing between languages. These toolkits focus on efficient memory representation, enabling zero-copy data transfer across languages. While highly optimized for data-centric workflows, they do not address the broader need for function-level interoperability. Kernel-FFI complements these systems by providing a general-purpose mechanism for invoking functions, creating objects, and managing state across languages.

\section{Conclusion}
\label{sec:conclusion}

In this paper, we presented Kernel-FFI, a novel approach to cross-language interoperability in interactive computing environments. By leveraging Jupyter kernels and introducing a flexible intermediate representation, Kernel-FFI enables seamless integration of multiple programming languages without requiring extensive boilerplate code or predefined schemas. The system supports not only basic function calls but also complex interactions involving object instantiation, custom types, and state management across language boundaries. Our evaluation demonstrates that Kernel-FFI successfully addresses the limitations of existing FFI solutions by providing transparent, language-agnostic cross-language function calls that are well-suited for the dynamic, interactive nature of modern notebook-based development workflows.

\bibliographystyle{ACM-Reference-Format}
\bibliography{main}

\end{document}